\documentclass[a4paper,fleqn,usenatbib,useAMS,onecolumn]{mnras}
\usepackage{geometry}                % See geometry.pdf to learn the layout options. There are lots.
\usepackage{amssymb}
\usepackage{amsmath}
\usepackage{graphicx}
\usepackage{epstopdf}

\usepackage{natbib}
\newcommand{\bs}{{\boldsymbol\xi}}
\newcommand{\be}{{\bf e}}

\newcommand{\bu}{{\bf u}}
\newcommand{\op}{{{\mathcal{L}}}}
\newcommand{\mB}{{{\mathcal{B}}}}

\newcommand{\bH}{{{\boldsymbol{\mathcal{H}}}}}
\newcommand{\bvar}{{{\boldsymbol{\varepsilon}}}}

\newcommand{\dU}{{\dot U}}

\newcommand{\siH}{{\mathcal H}}

\newcommand{\br}{{\bf r}}

\newcommand{\bT}{{\bf T}}
\newcommand{\bnabla}{{\boldsymbol\nabla}}
\newcommand{\bB}{{\bf B}}
\newcommand{\eye}{{\bf I}}
\newcommand{\rhat}{{\bf e}_r}

\title[Seismic sensitivity to Lorentz stresses]{Seismic sensitivity of Normal-mode Coupling to Lorentz stresses in the Sun}
\author[S. M. Hanasoge]{Shravan M. Hanasoge\thanks{Contact e-mail: \href{mailto:hanasoge@tifr.res.in}{hanasoge@tifr.res.in}}%\thanks{Department of Astronomy and Astrophysics, Tata Institute of Fundamental Research, Mumbai, India}
\\
Department of Astronomy and Astrophysics, Tata Institute of Fundamental Research, Mumbai, India}
\date{}

\pubyear{2017}

\begin{document}
\label{firstpage}
\pagerange{\pageref{firstpage}--\pageref{lastpage}}
\maketitle

\begin{abstract} 
Understanding the governing mechanism of solar magnetism remains an outstanding challenge in astrophysics.
Seismology is the most compelling technique with which to infer the internal properties of the Sun and stars. 
Waves in the Sun, nominally acoustic, are sensitive to the emergence and cyclical strengthening of magnetic field, evidenced by measured changes in resonant oscillation frequencies that are correlated with the solar cycle. The inference of internal Lorentz stresses from these measurements has the potential to significantly advance our appreciation of the dynamo. Indeed, seismological inverse theory for the Sun is well understood for perturbations in composition, thermal structure and flows but, is not fully developed for magnetism, owing to the complexity of the ideal magnetohydrodynamic (MHD) equation. Invoking first-Born perturbation theory to characterize departures from spherically symmetric hydrostatic models of the Sun and applying the notation of generalized spherical harmonics, we calculate sensitivity functions of seismic measurements to the general time-varying Lorentz stress tensor. 
We find that eigenstates of isotropic (i.e. acoustic only) background models are dominantly sensitive to isotropic deviations in the stress tensor and much more weakly so to anisotropic stresses (and therefore challenging to infer). The apple cannot fall far from the tree.
%With this theoretical development, the helioseismic inverse problem is now fully formally posed.
\end{abstract}


\begingroup
\let\clearpage\relax
%\tableofcontents
\endgroup
\newpage
\maketitle

\section{Introduction}
The cycling of the Sun's magnetic field, occurring on the time scale of approximately 11 years, causes luminosity changes and affects Earth's climate and space and geo-magnetic environments \citep{schrijver00,solanki2013}. Magnetism in the Sun is a multi-scale phenomenon, ranging from the system size ($R_\odot = 695,700$ km) to a few km. Understanding Lorentz stresses on large scales lends insight to the processes that drive the solar dynamo. Because the internal layers of the Sun are opaque to radiation and therefore inaccessible by optical imaging, seismology provides a unique and powerful technique with which to study the interior. A variety of seismic measurements in the Sun are used to infer its properties, such as global \citep[e.g.][]{jcd02} and local normal-mode frequencies \citep{hill88}, wave travel times \citep{duvall}, mode coupling \citep{woodard16} and holograms \citep{lindsey97}. 
For instance, at solar maxima when Lorentz stresses reach their peak magnitudes, solar normal mode frequencies are observed to be elevated in relation to their values at solar minima \citep[][]{woodard90}.
Using measured changes in frequencies or other seismic measurements to infer the internal state of the Sun is the goal of helioseismology.

Linear magnetohydrodynamics (MHD), the theory of small-amplitude wave propagation in magnetised media, is used to describe the physics of helioseismic oscillations \citep[e.g.][]{goedbloed2004, jcd_notes}. Acoustic waves are transformed to magnetosonic and incompressible Alfv\'{e}n waves, akin to vibrations of elastic media \citep[e.g.][]{roberts00}. Inviscid fluids only support pressure stresses, which act locally isotropically. In contrast, magnetic fields and flows cause waves to propagate anisotropically. { Alfv\'{e}n waves, which propagate only in the presence of magnetic fields, behave as vertically polarised shear waves, thereby adding to the anisotropy}.  { It is important to recognise that wave propagation in the limit of vanishingly small magnetic fields is not the same as in the zero-field case since Alfv\'{e}n waves exist is the former and cannot exist in the latter. Because the wavelength of Alfv\'{e}n waves scales linearly with magnetic field strength, it becomes infinitesimally small in the limit of vanishing field strength. In contrast, when the field strength is identically zero, waves do not disperse into magneto-acoustic and Alfv\'{e}n modes, and are described as purely acoustic oscillations.} Indeed, for this reason, MHD may act as a singular perturbation to an otherwise hydrostatic state although \citet{gizon06} showed that regular perturbation theory could be used to effectively predict changes in seismic measurements due to magnetic fields { when the ratio of magnetic-to-hydrostatic pressure is small}. However, owing to the tensor nature of the MHD equation, it has thus far not been possible to obtain a formal relationship between magnetic fields and attendant deviations in seismic measurements \citep[not for the want of trying, see e.g.][]{goughmag,goode05}. At the heart of the inverse problem is this relationship, i.e. the construction of sensitivity functions or kernels that capture the dependence of seismic measurements to perturbations in the solar model. 

It has been possible to obtain kernels for sound-speed and flow anomalies \citep[e.g.][]{jcd_notes,gizon02, birch, birch07, hanasoge11,boening, gizon17, mandal17} and for numerically computing small deviations around an existing magnetised state \citep{hanasoge11,hanasoge12_mag}. However without the theoretical machinery to account for the full anisotropy of the MHD equation, modelling the direct influence of general Lorentz stresses on seismic variables has eluded resolution thus far \citep[the impact of toroidal magnetic fields on normal mode frequencies was considered by][]{goughmag}. Prior approaches invoke assumptions on the field geometry to make the problem tractable, however at the cost of potentially diminishing inferential accuracy \citep{goughmag, goode05}.

Hydrodynamic pressure increases rapidly with depth in the Sun, implying that Lorentz stresses grow comparatively weaker, allowing for the application of perturbation theory. Whereas in near-surface layers, magnetic pressure is comparable to or greater than hydrodynamic pressure, and therefore surface magnetism represents a large deviation (e.g. sunspots). This latter problem deals with perturbing around a given model of a sunspot to fit seismic measurements and requires the application of iterative numerical methods \citep[][]{hanasoge11,hanasoge12_mag}. In contrast, the present technique allows for the direct inference of the Lorentz stress and treats it as a perturbation from a hydrostatic state. %The choice of strategy for the inverse problem thus depends on the strength of the magnetic feature in question.
Applying solid-Earth mode theory and treating field as a regular perturbation to the helioseismic wave equation, we derive the scattering matrix due to Lorentz stresses for mode coupling-measurements. Geophysical mode theory is particularly well suited to the problem at hand because it has been designed to address wave physics in the anisotropic Earth \citep[e.g.][]{DT98}. 

Resonant modes, which are computed for a given spherically symmetric structure model of the Sun, are nominally ``uncoupled" in that they are independent of one another. 
Deviations from this spherically symmetric state cause mode scattering, inducing correlations among different modes in the reference model and they become ``coupled". {  For temporally stationary perturbations to a given linear wave operator, mode scattering occurs at constant frequency. Here we allow the perturbation to vary in time and model the resultant coupling between modes} at different frequencies as well. The proximity of modes to one another, i.e. in terms of spatial and temporal frequencies, determines the extent of mode coupling. The closer the modes are, the stronger the scattering-induced correlation. Although we only outline the theory for mode-coupling measurements, the formalism here is immediately suitable to computing Lorentz kernels for normal-mode and travel-time measurements. 

\section{Helioseismic measurements}
For a non-rotating, non-magnetic, { undamped}, spherically symmetric model of the Sun, the linear acoustic wave equation for displacement $\bxi(\br,\omega)$ is given by \citep[e.g.][]{jcd_notes}
\begin{equation}
\op_0\bxi = -\rho\,\omega^2 \bxi -\bnabla(\rho c^2 \bnabla\cdot\bxi + \rho \bxi\cdot{\rhat} g)- g\rhat\bnabla\cdot(\rho\bxi) =0,\label{fullop}
\end{equation}
where $\omega$ is temporal frequency, $c(r)$ the sound speed, ${g}(r)$ is gravity, $\rho(r)$ the density and $\bnabla$ the covariant spatial derivative and { $\op_0$ is the unperturbed wave operator}. The eigenfrequencies and eigenfunctions of the Hermitian operator { $\op_0$ in equation}~(\ref{fullop}) are real.
We employ spherical coordinates with radius, colatitude and longitude denoted by $\br = (r,\theta,\phi)$ and unit vectors $(\rhat,\be_\theta,\be_\phi$) respectively. Non-radial variations in $\rho$, $c$, rotation, material circulations and magnetism are considered perturbations to the operator~(\ref{fullop}). For the analysis here, we assume that $\rho$ is only a function of radius.

A general wavefield $\bxi$ may be written in terms of mode eigenfunctions $\bs_k$ thus $\bxi = \sum_k a_k(\omega)\,\bs_k(\br),$ where $a_k$ denotes the contribution of mode $k$. Resonant modes are identified by quantum numbers $k=(\ell,m,n)$, where $\ell$ is spherical-harmonic degree, $m$ azimuthal order and $n$, radial order. Writing equation~(\ref{fullop}) in operator notation for the eigenfunction $\bs_k$ associated with mode $k$,
\begin{equation}
 \op_0\bs_k = \rho\,\omega_k^2\bs_k, \label{shorthand}
\end{equation}
where $\omega_k$ is the (real) resonant frequency. 
The eigenfunctions $\bs_k$ form an orthonormal basis when integrated over the solar volume $\odot$,
\begin{equation}
\int_\odot d\br\,\rho\,\bs^*_m\cdot\bs_n = \delta_{mn}.
\end{equation}
Modes are continuously randomly excited by near-surface convection in the Sun, resulting in stochastic time series' $a_k(t)$ for each mode $k$. 
For an unperturbed spherically symmetric solar model, we have $\langle a^{\omega'*}_j\, a^{\omega}_k\rangle = |R^\omega_k|^2 \delta(\omega - \omega')\, \delta_{jk}$ \citep[e.g.][]{woodard07}, where
\begin{equation}
R^\omega_{k} = \frac{1}{\bar\omega_{k}^2 - \omega^2},%\approx\frac{1}{2\omega_{k}(\omega_{k} - i\gamma_{k}/2 - \omega)},
\end{equation}
which only contributes at frequencies close to resonance. Note that the dependence on frequency is now expressed through a superscript to be consistent with \citet{woodard16} and \citet{hanasoge_etal_2017}. Solar modes experience a small degree of attenuation $\gamma_k\ll \omega_k$ that we take into account by perturbing only the eigenfrequency $\bar\omega_k \approx \omega_k- i\gamma_k/2$ in equation~(\ref{shorthand}), leaving the eigenfunction unchanged.  
Thus the cross-spectral measurement, $\langle a^{\omega'*}_j\, a^{\omega}_k\rangle$ when $j\ne k$, is non-zero only when solar structure departs from purely acoustic spherical symmetry. The Michelson Doppler Imager \citep{scherrer95} and Helioseismic and Magnetic Imager \citep{hmi} space missions, which have together observed some 20 years of the spherical-harmonic coefficients $a_k(t)$, allow us to measure these deviations.

Now consider a time-varying perturbation to the operator, $\delta\op_\omega$, which will in turn modify the wavefield by an amount $\delta\bxi$,  
\begin{eqnarray}
(-\rho\,\omega^2  + \op_0 +\delta\op_\omega)(\bxi + \delta\bxi) = 0.\label{eq.pert}
\end{eqnarray}
The subscript $\omega$ on $\delta\op$ denotes the frequency dependence of the perturbation (arising from its time variability).
The perturbed wavefield is written as a linear superposition of the original eigenfunctions,
\begin{equation}
\delta\bxi = \sum_{j} \delta a^\omega_j\,\bs_j.\label{recast}
\end{equation}
With some algebra \citep[e.g.][]{woodard16,hanasoge_etal_2017}, we arrive at a model for cross-spectral correlations,
\begin{equation}
\langle  a^{{\omega+\sigma}}_{k'}\,\delta a^{\omega*}_{k} + \delta a^{{\omega+\sigma}}_{k'}\, a^{\omega*}_{k} \rangle\approx H\Lambda^{k'}_{k}(\sigma),\label{pert2}
\end{equation}
where 
%\begin{equation}
%R_{k} = \frac{1}{\omega_{k}^2 - \omega^2}\approx\frac{1}{2\omega_{k}(\omega_{k} - i\gamma_{k}/2 - \omega)},
%\end{equation}
%is the mode power profile, 
$H= R^{\omega+\sigma}_{k'}\, |R^\omega_{k}|^2 + R^\omega_{k}\, |R^{(\omega+\sigma)*}_{k'}|^2$ and the coupling or scattering matrix $\Lambda$
\begin{equation}
\Lambda^{k'}_k(\sigma) = -\int_\odot d\br \,\,\bs^*_{k'}\cdot\delta\op_{\sigma}\,\bs_k,\label{coupmat}
\end{equation}
captures the extent of scattering, mediated by perturbation operator $\delta\op_\sigma$, from mode $k$ to $k'$. 
In equation~(\ref{coupmat}), $\delta\op_\sigma$ is the perturbation operator measured at temporal frequency channel $\sigma$. 
Because the Lorentz stress is a real quantity in the spatio-temporal domain and linear MHD is self-adjoint \citep[e.g.][]{goedbloed2004,hanasoge11}, we have $\Lambda^{k*}_{k'}(-\sigma) = \Lambda^{k'}_k(\sigma)$.
%Making use of terminology originally set out in \citet{lavely92}, we first introduce spherical harmonics $Y_\ell^m(\theta, \phi)$, the basis set on which eigenfunctions are projected. The generalised spherical harmonic $Y_\ell^{Nm} = D^\ell_{Nm}(\phi,\theta,0)$, where $D^\ell_{Nm}$ is the Wigner rotation matrix that relates spherical harmonics in rotated frames, will also play an important role in the analysis.

%The mode eigenfunction describing oscillations in a non-rotating, non-magnetized, spherically symmetric model of the Sun may be written so $\bs_k = U(r)\,Y^m_\ell\,\rhat + V(r)\,\bnabla_h Y^m_\ell$, where $k = (\ell, m, n)$ denotes a specific mode, and $\bnabla_h$ is the horizontal covariant derivative \citep[e.g.][]{lavely92, jcd_notes}. Rewriting the eigenfunction in terms of generalised spherical harmonics and simplifying (see Eq.~[\ref{recurse}] and Eq.~[\ref{specuse}]), we obtain,
A general time-varying magnetic field in spherical geometry is written thus
\begin{equation}
\bB(\br, \sigma) = \sum_{s=0}^\infty\sum_{t=-s}^s\left( u^t_s Y^{t}_s \rhat + v^t_s \bnabla Y^{t}_s\right) %\right.\label{generic}\\
+w^t_s \rhat\times  \bnabla Y^{t}_s,\label{vecmag}% = &&\left. + 
\end{equation}
where $Y^t_s$ are spherical harmonics of azimuthal order $t$ and spherical harmonic degree $s$, $u^t_s(r,\omega), v^t_s(r,\omega)$ constitute polodial-field coefficients and the $w^t_s(r,\omega)$ term represents toroidal field and $\omega$ is temporal frequency. The toroidal component by construction is solenoidal, i.e. $\bnabla\cdot(w^t_s \rhat\times  \bnabla Y^{s}_t) = 0$. In order to enforce $\bnabla\cdot\bB = 0$, the poloidal coefficients must obey $\partial_r(r^2 u^t_s) = s(s+1) r v^t_s$. 

Manipulating vectors and tensors in spherical geometry is simplified when using
generalised spherical harmonics \citep[][]{phinney73, DT98}. The generalised coordinate system is given by %$\be_0 = \be_r$ and
\begin{equation}
\be_0 = \be_r,\,\,\,\,\be_+ = -(\be_\theta + i\be_\phi)/\sqrt2,\,\,\,\,\be_- = (\be_\theta - i\be_\phi)/\sqrt2,\,\,\,\,\,\be^*_0 = \be_0,\,\,\,\,\,
\be^*_+ = - \be_-,\,\,\,\,\,\,\be^*_- = - \be_+.
\end{equation}
Eigenfunctions for an unperturbed spherically symmetric solar model may be expanded using spheroidal functions thus \citep[e.g.][]{jcd_notes,phinney73},
\begin{eqnarray}
&&\bs_k =  \sum_{\ell,m}u^m_\ell Y^{m}_\ell \rhat + v^m_\ell \bnabla Y^{m}_\ell \label{eigsun}\\
&&=\sum_{\ell,m} \xi^0_k Y^{0,m}_\ell \,\be_0+ \xi^-_k Y^{-1,m}_\ell\,\be_- + \xi^+_k Y^{1,m}_\ell\,\be_+,
\label{mode.eig}
\end{eqnarray}
where $Y^{Nm}_\ell$, which are generalised spherical harmonics, are related to elements of the Wigner rotation matrix  $Y^{Nm}_\ell= d^\ell_{Nm}(\theta,\phi)\,e^{im\phi}$ \citep[see Appendix D of][]{DT98}. Equation~(\ref{eigsun}) states that the simplest form of the solar eigenfunction comprises entirely spheroidal modes and lacks toroidal modes such as shear waves \citep[e.g. Chapter 8 of][]{DT98}, resulting in $\xi_k^+ = \xi_k^-$. 
%Since the basis vectors are complex, we make note of the following relationship for the conjugated eigenfunction
%\begin{equation}
%\bs^*_k =\sum_{s,t} \xi^{0*}_k Y^{0,m *}_\ell \,\be_0 - \xi^{-*}_k Y^{-1,m*}_\ell\,\be_+ - \xi^{+*}_k Y^{1,m*}_\ell\,\be_-.
%\label{mode.eigstar}
%\end{equation}
Equation~(\ref{vecmag}) for a general field is also rewritten using $\pm,0$ notation,
\begin{equation}
\bB(\br,\sigma) =\sum_{s,t}B^{0}_{st} Y^{0t}_{s} \be_0 + B^{+}_{st} Y^{1,t}_s \be_+ + B^{-}_{st} Y^{-1,t}_s \be_-,\label{pmmag}
\end{equation}
%where $u,v$ are poloidal and  $w$ are toroidal field coefficients respectively.
and the solenoidal condition on the field translates to
%\begin{equation}
$B_{st}^{+} + B_{st}^{-} = {\partial_r(r^2B_{st}^0)}/{r\Omega^s_0}$.
%\end{equation}
The $s,t$ indices occur as subscripts in equation~(\ref{pmmag}) for convenience.
%\newline%which unfortunately does not appear to translate to an equivalent constraint on the stress tensor. 
%which implies two coefficients, one poloidal and one toroidal, fully capture magnetic fields.\newline
%\newline
\section{MHD Equation} 
\begin{figure}%[t!]
\begin{center}
\includegraphics[width=\linewidth,clip=]{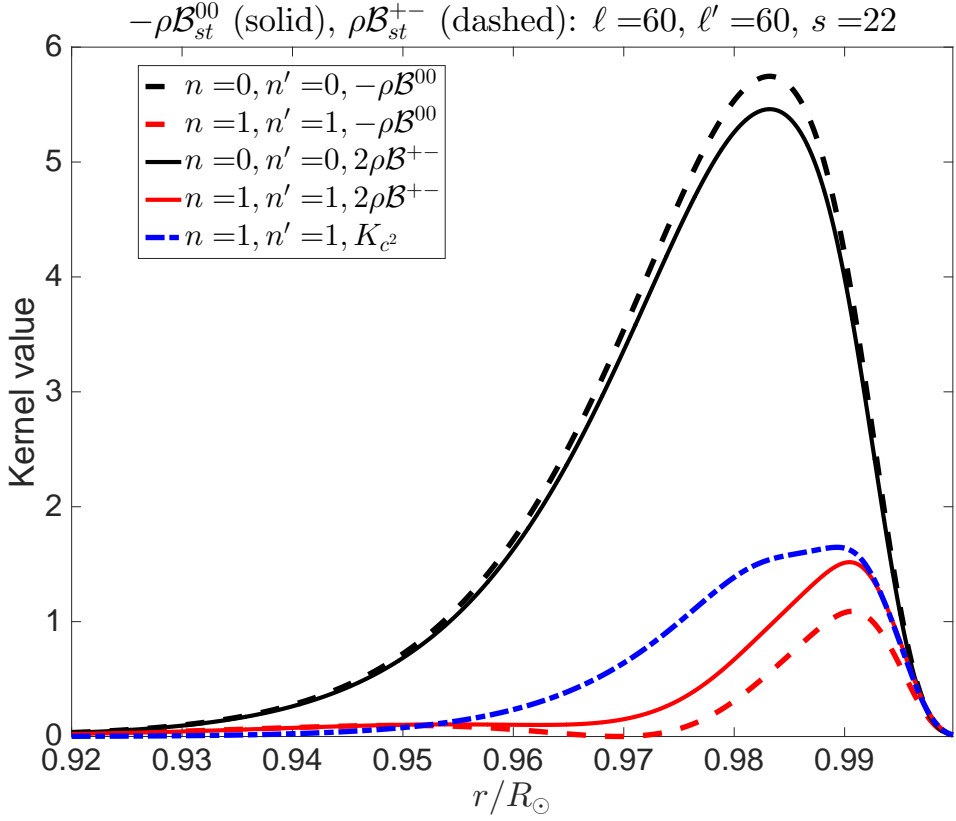}\vspace{-0.5cm}
\end{center}
\caption{The sensitivity of $\ell=60$ self-coupled normal modes to isotropic components $h^{00}$ ($\propto B_r B_r$, the radial magnetic energy) and $h^{+-}$ ($\propto B_\theta B_\theta + B_\phi B_\phi $, the transverse magnetic energy). { The expressions for the sensitivity kernels $\mB_{st}^{00}$ and $\mB_{st}^{+-}$ may be found in equations~(\ref{b00}) and~(\ref{bpm}) respectively. The dependence on $t$ is introduced through Wigner-3$j$ symbols linking $m, m'$ and $t$ in the expressions for kernels and only modify the overall sign and amplitude. We therefore ignore that term here (see Appendix~\ref{derive.kernels}).} The presence of magnetic fields cause modes to scatter, modifying the frequencies and amplitudes. Also plotted is the sound-speed kernel for comparison. The $f$ mode ($n=0$) is dominantly sensitive to the magnetic field whereas the $p_1$ mode has relatively greater sensitivity to variations in sound speed. 
%Kernels for coupling between modes of radial order $n$ have $n$ nodes in radius.
}
\label{field}
\end{figure}

\begin{figure}%[t!]
\begin{center}
\includegraphics[width=\linewidth,clip=]{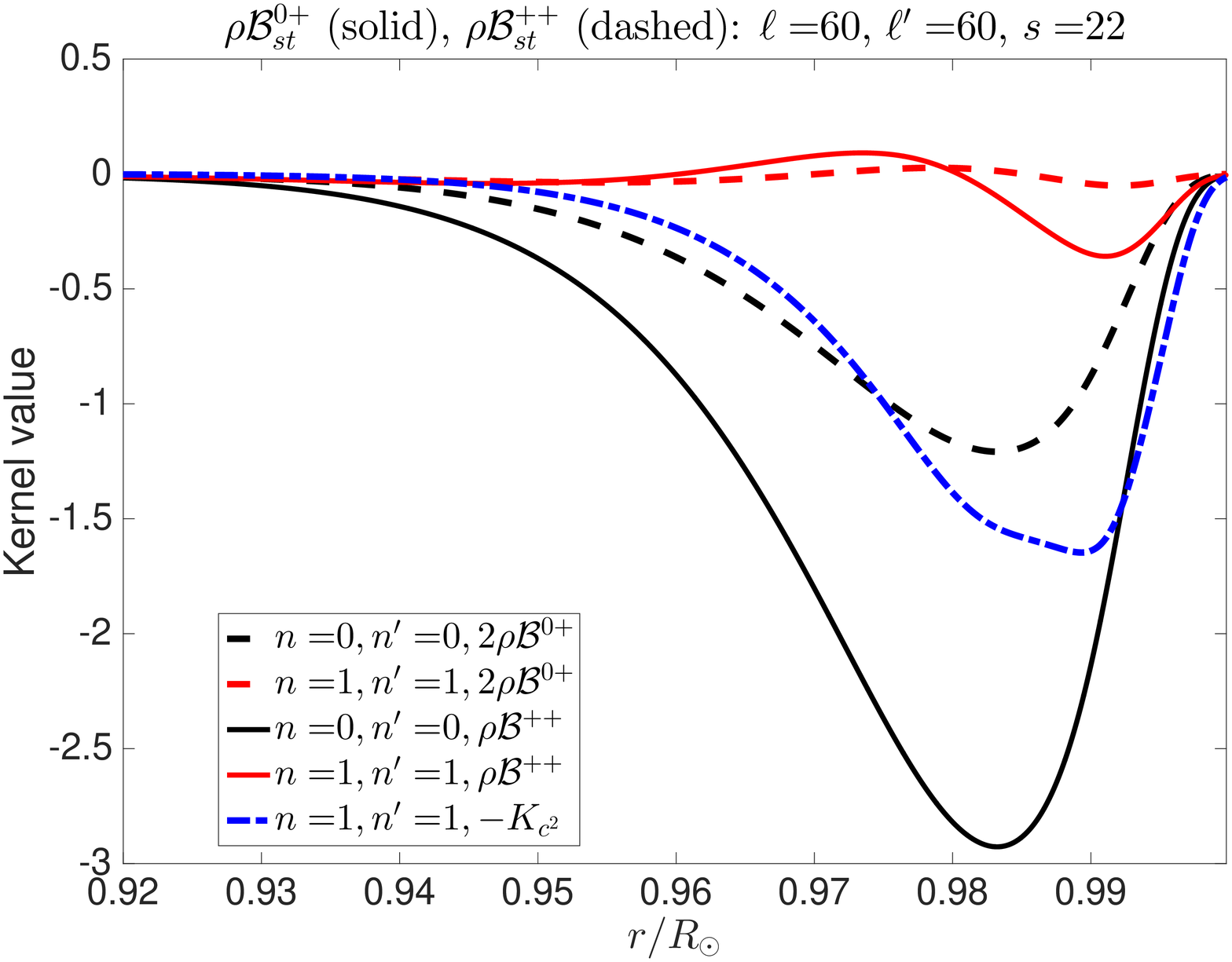}\vspace{-0.5cm}
\end{center}
\caption{The sensitivity of $\ell=60$ self-coupling to anisotropic stresses $h^{0+}$ ($\propto B_r B_\theta$ and $ B_r B_\phi$) and $h^{++}$ (relating to $ B_\theta B_\phi$). { The expressions for the sensitivity kernels $\mB_{st}^{0+}$ and $\mB_{st}^{++}$ may be found in equations~(\ref{b0p}) and~(\ref{bpp}) respectively. The dependence on $t$ is introduced through Wigner-3$j$ symbols linking $m, m'$ and $t$ in the expressions for kernels and only modify the overall sign and amplitude. We therefore ignore that term here (see Appendix~\ref{derive.kernels}).} The sensitivity to anisotropy is weaker than to isotropic stresses (compare with Figure~\ref{field}). Also plotted is the sound-speed kernel for $n = n' = 1$ for comparison. The $f$ mode ($n=0$) continues to be sensitive to the anisotropic components magnetic field whereas the $p_1$ mode is much more weakly sensitive (compare with the sound-speed kernel). %Wigner-3$j$ symbols in the expressions for the kernels codify the dependencies on $m, m'$ and $t$, only serving however to modifying the overall sign and amplitude. We therefore ignore that term here (see Appendix~\ref{derive.kernels}). 
%Kernels for coupling between modes of radial order $n$ have $n$ nodes in radius.
}
\label{field2}
\end{figure}

\begin{figure}%[t!]
\begin{center}
\includegraphics[width=\linewidth,clip=]{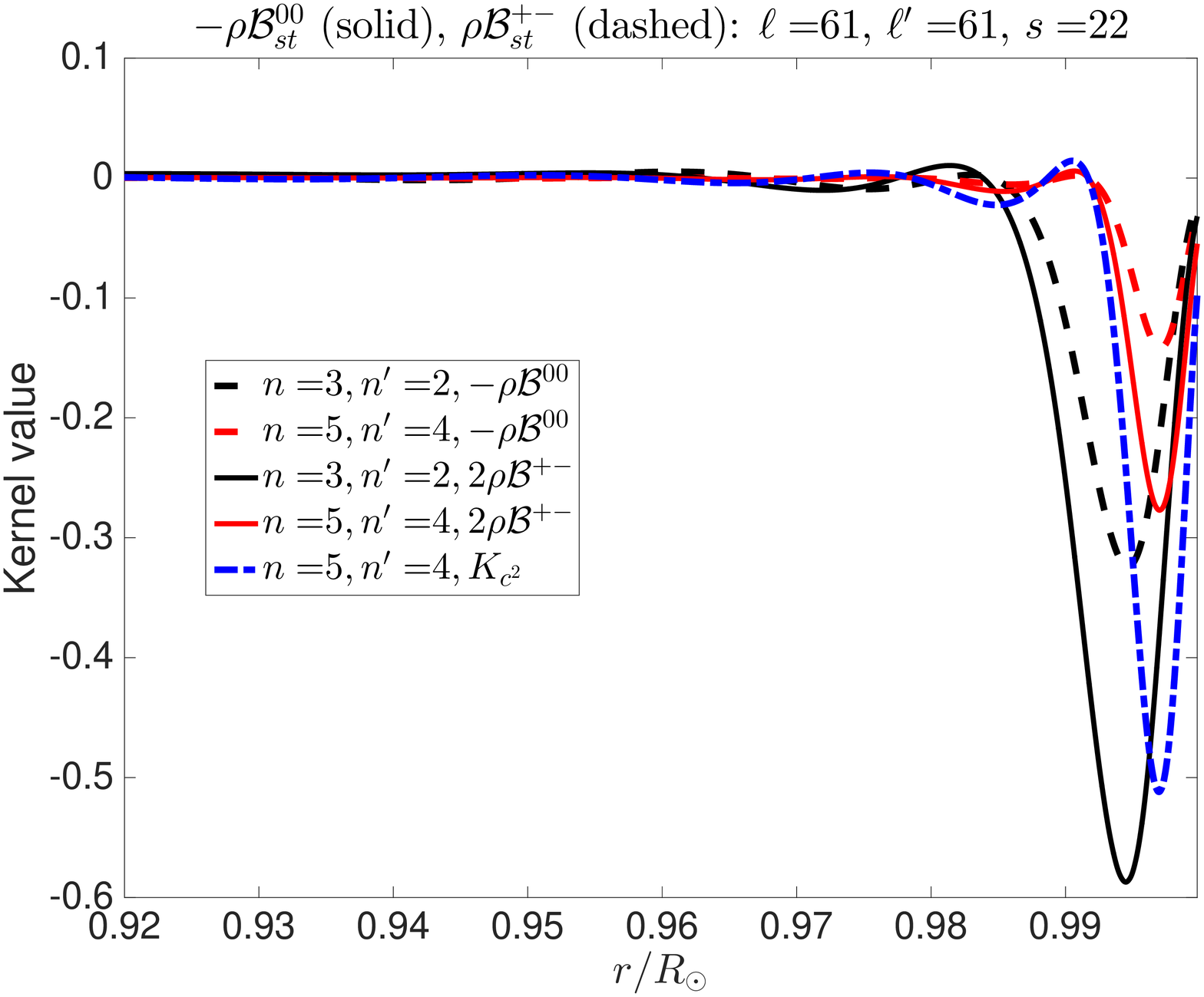}\vspace{-0.5cm}
\end{center}
\caption{The sensitivity of $\ell=61$ cross-spectrally coupled normal modes to isotropic components $h^{00}$ ($\propto B_r B_r$, the radial magnetic energy) and $h^{+-}$ ($\propto B_\theta B_\theta + B_\phi B_\phi$, the transverse magnetic energy). { The expressions for the sensitivity kernels $\mB_{st}^{00}$ and $\mB_{st}^{+-}$ may be found in equations~(\ref{b00}) and~(\ref{bpm}) respectively. The dependence on $t$ is introduced through Wigner-3$j$ symbols linking $m, m'$ and $t$ in the expressions for kernels and only modify the overall sign and amplitude. We therefore ignore that term here (see Appendix~\ref{derive.kernels}).} The sensitivity of cross coupling is weaker by an order of magnitude than self-coupled modes (compare with Figure~\ref{field}). Also plotted is the sound-speed kernel for $n = 5, n' = 4$ for comparison. While it is interesting to see that modes continue to be sensitive to magnetic field, the attendant seismic signatures are seen to be very similar to those of sound-speed perturbations. %Wigner-3$j$ symbols in the expressions for the kernels codify the dependencies on $m, m'$ and $t$, only serving however to modifying the overall sign and amplitude. We therefore ignore that term here (see Appendix~\ref{derive.kernels}). 
%Kernels for coupling between modes of radial order $n$ have $n$ nodes in radius. The $f$ mode ($n=0$) continues to be sensitive to the anisotropic components magnetic field whereas the $p_1$ mode is much more weakly sensitive (compare with the sound-speed kernel). 
}
\label{field3}
\end{figure}

\begin{figure}%[t!]
\begin{center}
\includegraphics[width=\linewidth,clip=]{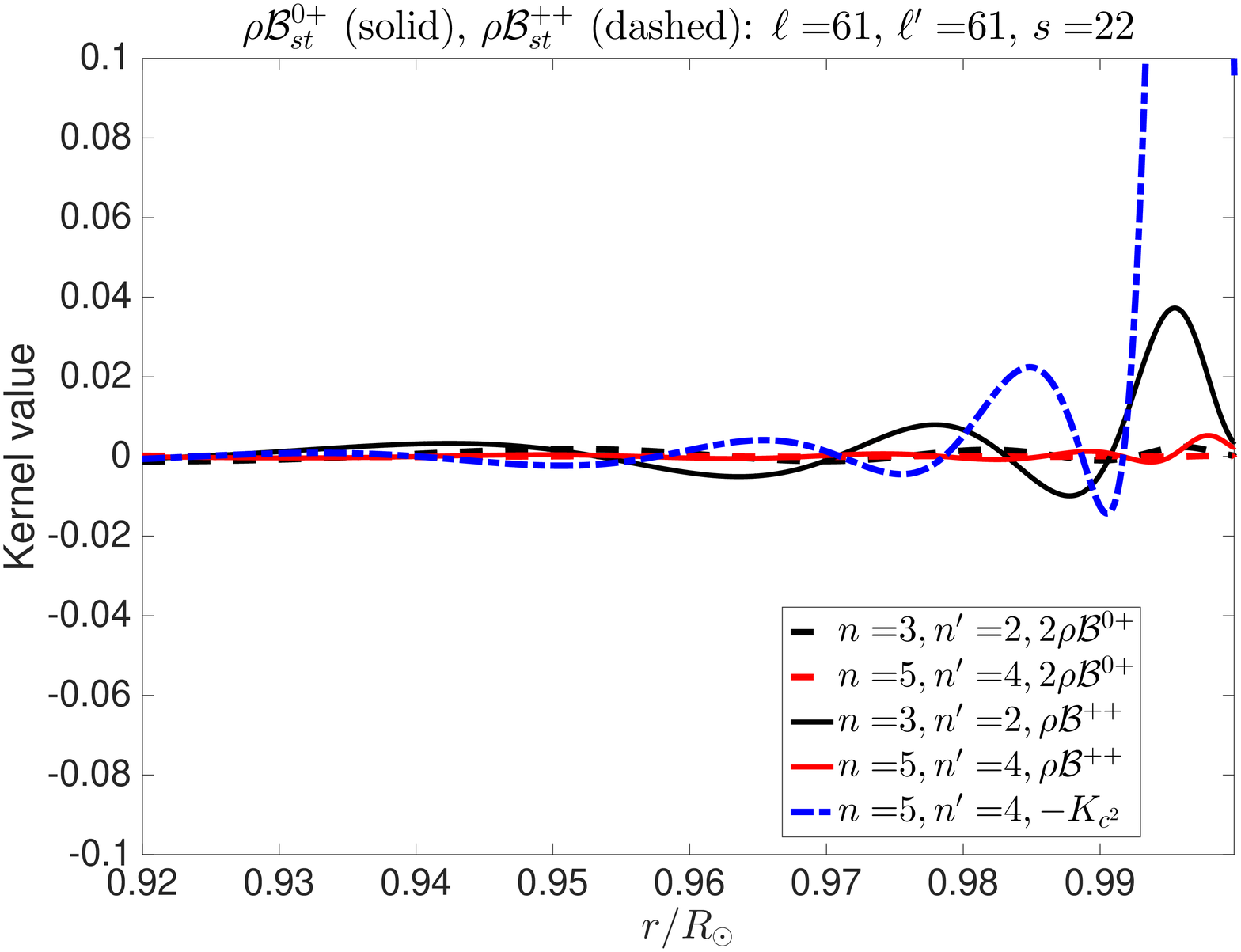}\vspace{-0.5cm}
\end{center}
\caption{The sensitivity of $\ell=61$ cross-spectral coupling to anisotropic stresses $h^{0+}$ ($\propto B_r B_\theta$ and $ B_r B_\phi$) and $h^{++}$ (relating to $ B_\theta B_\phi$). { The expressions for the sensitivity kernels $\mB_{st}^{0+}$ and $\mB_{st}^{++}$ may be found in equations~(\ref{b0p}) and~(\ref{bpp}) respectively. The dependence on $t$ is introduced through Wigner-3$j$ symbols linking $m, m'$ and $t$ in the expressions for kernels and only modify the overall sign and amplitude. We therefore ignore that term here (see Appendix~\ref{derive.kernels}).} The sensitivity to anisotropy is significantly weaker than to isotropic stresses (compare with Figures~\ref{field3} and~\ref{field4}). Also plotted is the sound-speed kernel for $n = 5, n' = 4$ for comparison, which is seen to be much larger in magnitude. %Wigner-3$j$ symbols in the expressions for the kernels codify the dependencies on $m, m'$ and $t$, only serving however to modifying the overall sign and amplitude. We therefore ignore that term here (see Appendix~\ref{derive.kernels}). 
%Kernels for coupling between modes of radial order $n$ have $n$ nodes in radius.
}
\label{field4}
\end{figure}

The action of magnetism is described using linearized ideal MHD, a model of small-amplitude fluctuations about an equilibrium \citep{goedbloed2004}. The time-varying Lorentz-stress tensor $\bH = \bB\bB$, where $\bB$ is the field, perturbs operator~(\ref{fullop}) thus,
\begin{equation}
\delta\op\bxi = -\bnabla\cdot\left[ \bH\cdot\bnabla\bxi +(\bnabla\bxi)^T\cdot\bH -\bxi\cdot\bnabla\bH%\right.\nonumber\\  &&\left.
 -2\bH\bnabla\cdot\bxi + (\bH:\eye\bnabla\cdot\bxi - \bH:\bnabla\bxi) \eye %\right. \nonumber\\ && \left.+ 
\bxi\cdot\bnabla \frac{\bH:\eye}{2}\, \eye\right].\label{pertop}
\end{equation}
We outline the derivation of equation~(\ref{pertop}) in Appendix~\ref{mhd.eqs}.
The dependence of $\bH$ on $\omega$ is not explicitly stated to reduce notational burden. Denoting the strain tensor $\bvar_k = [\bnabla\bs_k + (\bnabla\bs_k)^T]/2$ and the unit dyad by $\eye$, i.e. $(\eye)_{ij} = \delta_{ij}$, the coupling coefficient linking two modes $k = (\ell, m, n)$ and $k' = (\ell', m', n')$ is 
\begin{equation}
\Lambda^{k'}_k = \int_\odot d\br\,\bH:\left[\bnabla\bs_k\cdot\bvar^*_{k'} + \bvar^*_{k'}\cdot(\bnabla\bs_k)^T \label{matrix} %\right.\\
- \bvar^*_{k'}\,\bnabla\cdot\bs_k- \bvar_{k}\bnabla\cdot\bs^*_{k'} + \eye\frac{\bnabla\cdot\bs_k\,\bnabla\cdot \bs^*_{k'}}{2}\right],%\nonumber
\end{equation}
where $(\bnabla \bs_k)_{ij} = \partial_i \xi_{k,j}$ and $(\bnabla \bs_k)^T_{ij} = \partial_j \xi_{k,i}$.
Using generalized spherical harmonics, we may expand $\bH(\br,\sigma)$, where $\sigma$ is temporal frequency, thus
\begin{equation}
\bH(\br,\sigma) = \sum_{i,j}\sum_{s,t} h^{ij}_{st}(r,\sigma)\, Y^{i+j,t}_s\, \be_i \,\be_j ,
\end{equation} %, although the solenoidal constraint on the field will further reduce this to 3
where $s$ is spherical harmonic degree, $t$ is azimuthal order, $h_{st}^{ij}(r,\sigma)$ is the $s,t$ coefficient of the $i,j$ component of the tensor $\bH$, and  $i$, $j$ are $\pm$ or  $0$. Because $\bH$ is symmetric and real in the spatio-temporal domain, the following relationships hold 
\begin{equation}
h_{st}^{0+} = h_{st}^{+0},\,\,\,\,\,\, h_{st}^{0-} = h_{st}^{-0},\,\,\,\,\,\, h_{st}^{-+} = h_{st}^{+-},\,\,\,\,\,\,(-1)^t\,h^{ij}_{st}(r,-\sigma) = [h^{ij}_{s,-t}(r,\sigma)]^*.\label{syms}\end{equation}
Thus $\bH$ only has 6 independent components and we use $(h_{st}^{++},h_{st}^{+0},h_{st}^{00},h_{st}^{+-},h_{st}^{-0},h_{st}^{--})$ to represent the tensor. %and the following symmetries apply, $h^{0+} = h^{+0}$, $h^{0-} = h^{-0}$ and $h^{-+} = h^{+-}$. 
Note that the inverse problem is for Lorentz stresses and not the field itself. The solenoidal condition on magnetic field could not readily be translated to an equivalent constraint on the Lorentz stress. We therefore do not incorporate it in the present analysis.
After tedious algebra (see Appendices~\ref{tensor.manip} and~\ref{derive.kernels}), we obtain the following { relation}
\begin{equation}
\Lambda^{k'}_k=\sum_{s,t}\int_\odot dr\, \mB_{st}^{00} h_{st}^{00} + \mB_{st}^{++} [h_{st}^{--}\,(-1)^{\ell'+\ell+s} + h_{st}^{++} ]%   \nonumber\\
+ 2\mB_{st}^{0+} [h_{st}^{0-}\,(-1)^{\ell'+\ell+s} + h_{st}^{0+}] + 2\mB_{st}^{+-} h_{st}^{+-} ,\label{invprob}
\end{equation}
%where $\kerne$ are defined in the SM.
where $\mB$, defined in Appendix~\ref{derive.kernels}, denote kernels for different components of the stress tensor, $\ell$ and $\ell'$ are the harmonic degrees associated with modes $k$ and $k'$ respectively that have become coupled due to Lorentz stresses. Using Wigner-3$j$ rules, integration over the 3-D sphere have been simplified to a 1-D integral over radius. Kernels $\mB^{00}$ and $\mB^{+-}$, whose superscripts sum to zero, capture the seismic sensitivity to isotropic, on-diagonal components of the stress tensor, the radial and transverse magnetic energies respectively. Kernels $\mB^{0+}$ and $\mB^{++}$ represent the sensitivity to off-diagonal, anisotropic Lorentz stresses.
We show examples of $\mB^{+-}$ and $\mB^{00}$ kernels in Figure~\ref{field} and $\mB^{++}$ and $\mB^{0+}$ kernels in Figure~\ref{field2}. { These are for self-coupled modes $\ell= \ell'$ and $n=n'$. In Figures~\ref{field3} and~\ref{field4}, we show cross-coupled kernels $\ell = \ell'$ and $n\neq n'$. In general we find that coupled modes are significantly more sensitive to isotropic components of the Lorentz stress tensor than the anisotropic terms.} The following relations connect stresses in real space to the intermediate $\pm,0$ variables,
\begin{eqnarray}
&& B_r B_r(\br,\sigma)  = \sum_{s,t} h_{st}^{00} Y^{0,t}_s,\label{relsbb} \\
&& B_r B_\theta(\br,\sigma)  = \sum_{s,t} \frac{h_{st}^{0-} Y^{-1,t}_s - h_{st}^{0+} Y^{1,t}_s}{\sqrt2},\nonumber\\
&& B_r B_\phi(\br,\sigma)  = -i\sum_{s,t} \frac{h_{st}^{0-} Y^{-1,t}_s + h_{st}^{0+} Y^{1,t}_s}{\sqrt2},\nonumber\\
&& B_\theta B_\theta(\br,\sigma) = \sum_{s,t}\frac{ h_{st}^{++} Y^{2,t}_s -2h_{st}^{+-} Y^{0,t}_s + h_{st}^{--} Y^{-2,t}_s}{2},\nonumber\\
&& B_\theta B_\phi(\br,\sigma)  = i\sum_{s,t} \frac{h_{st}^{++} Y^{2,t}_s - h_{st}^{--} Y^{-2,t}_s}{2},\nonumber\\
&& B_\phi B_\phi(\br,\sigma)  = \sum_{s,t}\frac{-h_{st}^{++} Y^{2,t}_s  - 2h_{st}^{+-} Y^{0,t}_s - h_{st}^{--} Y^{-2,t}_s}{2} .\nonumber
\end{eqnarray}
The inverse problem~(\ref{invprob}) indicates that modes with even $\ell + \ell' + s$ are only sensitive to $h^{00}$, $h^{+-}$, $h^{--} + h^{++}$, $h^{0+} + h^{0-}$ those with odd $\ell + \ell' + s$ only sense $h^{--} - h^{++}$ and $h^{0+} - h^{0-}$. This is also encountered when imaging flows for instance, where kernels with odd $\ell + \ell' + s$ are sensitive only to toroidal flows and kernels with even $\ell + \ell' + s$ are sensitive only to poloidal flows \citep[Appendices~\ref{gentoreal} and~\ref{flowkern} of this article and Appendices C and D of][]{lavely92}.

\section{Discussion} The foregoing analysis brings to light a technique to elegantly compute the influence of general anisotropic magnetic stresses on seismic variables. The results pave the way for formally inferring the Lorentz stress tensor using seismic measurements \citep[e.g.][]{goughmag,goode05}. We find that modes are much more sensitive to the diagonal components of the tensor, i.e. the magnetic energies, than off-diagonal, anisotropic terms.
The kernels appear to naturally separate out regular and singular perturbations associated with magnetic fields. The component of the Lorentz stress that is a regular perturbation is isotropic, behaving as sound-speed anomalies might, as demonstrated by the kernels in Figure~\ref{field}.
Seismic measurements are primarily sensitive to radial and transverse magnetic energies, which are the diagonal components the stress tensor. In contrast, the anisotropic behaviour of the magnetic field represents a singular perturbation to the original model, since the background does not contain anisotropy { (e.g. adding rotation or magnetic fields to it could induce anisotropy)}. A direct manifestation of anisotropy is the appearance of Alfv\'{e}n waves, which is not permitted in hydrodynamics. Modes computed around a hydrostatic background are far less sensitive to these stresses, as Figure~\ref{field2} suggests.
%%That the limit of arbitrarily small field (where Alfv\'{e}n waves exist) is not the same as the zero-field case is the clearest indication that this is an example of a singular perturbation. 
Indeed, deviations from models are primarily of the same character as the model itself and isotropy only begets deviations of the isotropic kind. 

{In appendix~\ref{flowkern}, we derive flow kernels using generalized spherical harmonics. However, in deriving these sensitivities, we ignore the second-order flow term, i.e. that goes as $\bnabla\cdot(\rho\,{\bf u} {\bf u}\cdot\bnabla\bxi)$ where ${\bf u}$ is the flow velocity. This term possesses some similarities with magnetic perturbations in that it takes the form ${\bf u}{\bf u}$ and flows obey the continuity condition $\bnabla\cdot (\rho {\bf u}) = 0$, akin to the divergence-free condition on magnetic fields. The second-order flow perturbation to the wave equation couples eigenfunctions according to $\rho({\bf u}\cdot\bnabla\bs_k)\cdot({\bf u}\cdot\bnabla\bs^*_{k'})$, taking on mathematical structure somewhat different from the magnetic terms of appendix~\ref{mhd.eqs}. While it may therefore be possible to distinguish between the coupling effects of the two, the data likely will not support discerning such subtleties owing to systematic effects such as spatial and temporal leakage. Moreover, inferring sub-surface magnetic fields will be difficult given that magnetic fields at the surface, where the magnetic-to-gas pressure is much greater than unity, couple modes strongly \citep[e.g.][]{goode05}. In particular, sunspots have locally very strong fields and the assumption of linearity between the perturbation and corresponding deviation in the measurement may break down. This implies that inferring Lorentz stresses in the sub-surface is a challenging problem. }

The present technique may also be used to model instantaneous and classical travel-time and amplitude measurements. First-Born scattering theory relies on Green's functions for computing kernels, and since Green's functions for spherically symmetric models may be expressed using equation~(\ref{eigsun}), the same vector harmonic basis as the eigenfunctions, the analysis proceeds unchanged. To describe normal-mode-frequency sensitivity, we may use self-coupling kernels as in Figures~\ref{field} and~\ref{field2}.

\section*{Acknowledgments}
SMH is grateful to David Al-Attar for a most useful conversation that set the analysis in motion. He also acknowledges support from Ramanujan fellowship SB/S2/RJN-73, the Max-Planck partner group program and thanks NYUAD's Center for Space Science. Jishnu Bhattacharya helped greatly by re-calculating and verifying these expressions using Mathematica.
%\end{acknowledgments}
\bibliographystyle{mnras}
\bibliography{ms}%/Users/Shravan/Dropbox
%
%  suggest that purely acoustic measurements only contain information about radial and transverse magnetic energies. and that building the entire Lorentz stress tensor would require recording Alfv\'{e}n wave.
\appendix
\section{MHD perturbation}\label{mhd.eqs}
 Linearized ideal MHD is a model of small amplitude fluctuations about an equilibrium \citep{goedbloed2004}.
The perturbation to the operator due to the presence of magnetism is given by
\begin{equation}
\delta\op = -\bnabla\cdot\left[ \bB\bB\cdot\bnabla\bxi +\bB\cdot\bnabla\bxi\,\bB -2\bB\bB\bnabla\cdot\bxi -(\bxi\cdot\bnabla\bB)\bB -\bB(\bxi\cdot\bnabla\bB) + %\right. \nonumber\\ &&\left. +
 B^2\bnabla\cdot\bxi\, \eye - \bB\bB:\bnabla\bxi\, \eye + \bxi\cdot\bnabla \frac{B^2}{2}\, \eye\right],
\end{equation}
where the notation ${\bf a} : {\bf b} = \sum_{i,j} a_{ij} b_{ji}$.
The term $\bnabla\cdot[(\bxi\cdot\bnabla\bB)\bB +\bB(\bxi\cdot\bnabla\bB)]$ in Einstein-index notation is $\partial_i[\xi_j\partial_j(B_i) B_k] + \partial_i(B_i\xi_j\partial_j B_k) = \partial_i[\xi_j\partial_j (B_i B_k)]$, i.e. $\bnabla\cdot[\bxi\cdot\bnabla(\bB\bB)]$. Writing the Lorentz-stress tensor as $\bH = \bB\bB$, the quantity of central interest here, we may rewrite the perturbation operator as
\begin{equation}
\delta\op\bxi = -\bnabla\cdot\left[ \bH\cdot\bnabla\bxi +(\bnabla\bxi)^T\cdot\bH -2\bH\bnabla\cdot\bxi -\bxi\cdot\bnabla\bH%\right.\nonumber\\ \left. + 
+\bH:\eye\bnabla\cdot\bxi\, \eye - \bH:\bnabla\bxi\, \eye + \bxi\cdot\bnabla \frac{\bH:\eye}{2}\, \eye\right].%\label{pertop}
\end{equation}
%The coupling matrix between two eigenfunctions mediated by the perturbation operator $\delta\op$ is given by
%\begin{equation}
%-\Lambda^{k'}_k = \int_\odot d\br\, \bs^*_{k'}\cdot\delta\op\,\bs_k .
%\label{coup.int}
%\end{equation}
We consider the coupling integral~(\ref{coupmat}) along with the definition of the operator~(\ref{pertop}) term by term,
\begin{equation}
-\int_\odot d\br\, \bs^*_{k'}\cdot\bnabla\cdot(\bH\cdot\bnabla\bs_k) = -\int_\odot d\br\, \bnabla\cdot[\bH\cdot(\bnabla\bs_k)\cdot\bs^*_{k'}] +  \int_\odot d\br\, \bH:[(\bnabla\bs_k)\cdot(\bnabla\bs^*_{k'})^T],
\end{equation}
\begin{equation}
-\int_\odot d\br\, \bs^*_{k'}\cdot\bnabla\cdot[(\bnabla\bs_k)^T\cdot\bH] = -\int_\odot d\br\, \bnabla\cdot[(\bnabla\bs_k)^T\cdot\bH\cdot\bs^*_{k'}] +  \int_\odot d\br\, \bH:[(\bnabla\bs_k)\cdot(\bnabla\bs^*_{k'})],
\end{equation}
\begin{equation}
2\int_\odot d\br\, \bs^*_{k'}\cdot\bnabla\cdot(\bH\bnabla\cdot\bs_k) = 2\int_\odot d\br\, \bnabla\cdot(\bH\cdot\bs^*_{k'}\,\bnabla\cdot\bs_k)  -2\int_\odot d\br\,\bH:(\bnabla\bs^*_{k'})\, \bnabla\cdot\bs_k,
\end{equation}
\begin{eqnarray}
&&\int_\odot d\br\, \bs^*_{k'}\cdot\bnabla\cdot(\bs_k\cdot\bnabla\bH) = \int_\odot d\br\, \bnabla\cdot[\bs_k\cdot(\bnabla\bH)\cdot\bs^*_{k'}] - \int_\odot d\br\, \bs_k\cdot(\bnabla\bH):(\bnabla\bs^*_{k'}),\nonumber\\
&&- \int_\odot d\br\, \bs_k\cdot(\bnabla\bH):(\bnabla\bs^*_{k'}) = - \int_\odot d\br\, \bnabla\cdot [\bs_k\bH:(\bnabla\bs^*_{k'})]
+\int_\odot d\br\, \bH:(\bnabla\bs^*_{k'})\bnabla\cdot\bs_k,\nonumber\\
\end{eqnarray}
\begin{equation}
-\int_\odot d\br\, \bs^*_{k'}\cdot\bnabla\cdot(\eye\bnabla\cdot\bs_k\bH:\eye) = -\int_\odot d\br\, \bnabla\cdot(\bs^*_{k'}\,\bnabla\cdot\bs_k\bH:\eye) + \int_\odot d\br\, \bH:(\eye\,\bnabla\cdot\bs^*_{k'}\,\bnabla\cdot\bs_k),
\end{equation}
\begin{equation}
\int_\odot d\br\, \bs^*_{k'}\cdot\bnabla\cdot(\eye\,\bH:\bnabla\bs_k) = \int_\odot d\br\, \bnabla\cdot(\bs^*_{k'}\,\bH:\bnabla\bs_k) - \int_\odot d\br\, \bH:[(\bnabla\bs_k)\,\bnabla\cdot\bs^*_{k'}],
\end{equation}
\begin{eqnarray}
&&-\frac{1}{2}\int_\odot d\br\, \bs^*_{k'}\cdot\bnabla\cdot[\eye\,\bs_k\cdot\bnabla(\bH:\eye)] =  \nonumber\\
&&-\frac{1}{2}\int_\odot d\br\, \bnabla\cdot[\bs^*_{k'}\,\bs_k\cdot\bnabla(\bH:\eye)]+ \frac{1}{2}\int_\odot d\br\, \bs_k\cdot\bnabla(\bH:\eye)\,\bnabla\cdot \bs^*_{k'},\\
&&\frac{1}{2}\int_\odot d\br\, \bs_k\cdot\bnabla(\bH:\eye)\,\bnabla\cdot \bs^*_{k'} = \frac{1}{2}\int_\odot d\br\, \bnabla\cdot(\bs_k\,\bnabla\cdot \bs^*_{k'}\,\bH:\eye) - \frac{1}{2}\int_\odot d\br\, \bH:(\eye\,\bnabla\cdot\bs_k\,\bnabla\cdot \bs^*_{k'}).\nonumber
\end{eqnarray}
The boundary contributions are assumed to vanish, allowing us to write the full coupling integral as
\begin{equation}
\Lambda^{k'}_k = \int_\odot d\br\,\bH:\left\{(\bnabla\bs_k)\cdot[(\bnabla\bs^*_{k'})^T + \bnabla\bs^*_{k'}] - (\bnabla\bs^*_{k'})\bnabla\cdot\bs_k - (\bnabla\bs_{k})\bnabla\cdot\bs^*_{k'} + \frac{1}{2}\eye\,\bnabla\cdot\bs_k\,\bnabla\cdot \bs^*_{k'}\right\}.
\end{equation}
Since $\bH$ and $\varepsilon$ are symmetric tensors, we may reduce the expression further
\begin{equation}
\Lambda^{k'}_k = \int_\odot d\br\,\bH:\left\{\bnabla\bxi_k\cdot\bvar^*_{k'} + \bvar^*_{k'}\cdot(\bnabla\bxi_k)^T - \bvar^*_{k'}\,\bnabla\cdot\bs_k - \bvar_{k}\bnabla\cdot\bs^*_{k'} + \frac{1}{2}\eye\,\bnabla\cdot\bs_k\,\bnabla\cdot \bs^*_{k'}\right\}\label{coupintfinal},
%= \int_\odot d\br\,\bH:\left\{\bvar_k\cdot[\bvar^*_{k'} - \eye\,\bnabla\cdot\bs^*_{k'}] + \bvar^*_{k'}\cdot[\bvar_k - \eye\bnabla\cdot\bs_k] + \frac{1}{2}\eye\,\bnabla\cdot\bs_k\,\bnabla\cdot \bs^*_{k'}\right\} \label{coupintfinal},
\end{equation}
where $\varepsilon_k = [\bnabla \bs_k + (\bnabla \bs_k)^T]/2$, and $(\bnabla\bs_k)_{ij} = \partial_i \xi_{k,j}$, $(\bnabla\bs_k)^T_{ij} = \partial_j \xi_{k,i}$. 

%We define the tensor $\mW$ as the expression within the parentheses on the right side of the double contraction of equation~(\ref{coupintfinal}),
%\begin{equation}
%\mW = \bnabla\bs_k\cdot\bvar^*_{k'} + \bvar^*_{k'}\cdot(\bnabla\bs_k)^T - \bvar^*_{k'}\,\bnabla\cdot\bs_k - \bvar_{k}\bnabla\cdot\bs^*_{k'} + \frac{1}{2}\eye\,\bnabla\cdot\bs_k\,\bnabla\cdot \bs^*_{k'}.\label{defw}
%\end{equation}

\section{Tensor manipulation}\label{tensor.manip}

Manipulating vectors and tensors in spherical geometry is simplified when using
generalised spherical harmonics \citep[][]{phinney73, DT98}. The generalised coordinate system is given by %$\be_0 = \be_r$ and
\begin{equation}
\be_0 = \be_r,\,\,\,\,\be_+ = -(\be_\theta + i\be_\phi)/\sqrt2,\,\,\,\,\be_- = (\be_\theta - i\be_\phi)/\sqrt2,\,\,\,\,\,\be^*_0 = \be_0,\,\,\,\,\,
\be^*_+ = - \be_-,\,\,\,\,\,\,\be^*_- = - \be_+.
\label{unitvec}
\end{equation}
and we have $\be_i\cdot\be_j =0$ with the exception of $\be_0\cdot\be_0 = 1$, $\be_+\cdot\be_- = - 1$. The following relations are also relevant, $\be_+\cdot\be^*_+ = 1 = \be_-\cdot\be^*_-$.
Using the rules and terminology of covariant differentiation developed by \citet{phinney73}
and defining the tensor $\bT_k = \bnabla\bs_k$, we obtain %equation~(\ref{Tsimple}).
\begin{eqnarray}
\bT_k = \sum_{\ell = 0}^\infty\sum_{m=-\ell}^\ell \sum_{\alpha,\beta} T_k^{\alpha\beta} Y^{\alpha+\beta, m}_\ell \be_\alpha \be_\beta,\\
T_k^{--} = \xi_k^{-|-} = \frac{1}{r}\Omega^\ell_{2}\,U^{-,m}_\ell ,\nonumber\\
T_k^{0-} = \xi_k^{-|0} = \dU^{-,m}_\ell,\nonumber\\
T_k^{+-} = \xi_k^{-|+} = \frac{1}{r}[\Omega^\ell_0\,U^{-,m}_\ell - U^{0,m}_\ell],\nonumber\\
T_k^{-0} = \xi_k^{0|-} = \frac{1}{r}[\Omega^\ell_0\,U^{0,m}_\ell - U^{-,m}_\ell],\nonumber\\
T_k^{00} = \xi_k^{0|0} = \dU^{0,m}_\ell,\nonumber\\
T_k^{+0} = \xi_k^{0|+} = \frac{1}{r}[\Omega^\ell_{0}\,U^{0,m}_\ell - U^{+,m}_\ell],\nonumber\\
T_k^{-+} = \xi_k^{+|-} = \frac{1}{r}[\Omega^\ell_0\,U^{+,m}_\ell - U^{0,m}_\ell],\nonumber\\
T_k^{0+} = \xi_k^{+|0} = \dU^{+,m}_\ell,\nonumber\\
T_k^{++} = \xi_k^{+|+} = \frac{1}{r}\Omega^\ell_{2}\,U^{+,m}_\ell(r).
\end{eqnarray}
{ The symbol $\xi_k^{a|b}$ denotes the derivative of $\xi_k^{a}$ with respect to the $b$ coordinate. The terms $\xi^{0,\pm}_k$ and $U, V$ are described in appendix~\ref{gentoreal} and coefficients $\Omega^\ell_2$ and $\Omega^\ell_0$ are defined in equations~(\ref{omegdef}) and~(\ref{omegequal}).}
Owing to the degeneracy between the $\pm$ components of the eigenfunction (Eq.~[\ref{unitvec}]), we have the following equivalences, $T_k^{--} = T_k^{++}$, $T_k^{+0} = T_k^{-0}$, $T_k^{-+} = T_k^{+-}$ and $T_k^{0+} = T_k^{0-}$.
The trace of this tensor is given by 
\begin{equation}
Tr(T_k) = Tr(\varepsilon_k) =  T_k^{00} - T_k^{-+} - T_k^{+-} = T_k^{00} - 2 T_k^{+-}%\nonumber\\
= \left\{\dU^{0,m}_\ell - \frac{1}{r}[\Omega^\ell_0\,(U^{-,m}_\ell + U^{+,m}_\ell) - 2U^{0,m}_\ell]\right\},
\end{equation}
where $\alpha, \beta$ take on the values $0, \pm1$. The symmetric strain tensor $\bvar_k = [(\bnabla\bs_k)^T + \bnabla\bs_k]/2$ is given by
\begin{eqnarray}
\bvar_k = \sum_{\ell = 0}^\infty\sum_{m=-\ell}^\ell \sum_{\alpha,\beta} \varepsilon_k^{\alpha\beta} Y^{\alpha+\beta, m}_\ell \be_\alpha \be_\beta,\\
\varepsilon_k^{--} = \xi_k^{-|-} = \frac{1}{r}\Omega^\ell_{2}\,U^{-,m}_\ell,\nonumber\\
\varepsilon_k^{-0} = \varepsilon_k^{0-}= \frac{\xi_k^{-|0} + \xi_k^{0|-}}{2} = \frac{1}{2}\{\dU^{-,m}_\ell + \frac{1}{r}[\Omega^\ell_0\,U^{0,m}_\ell - U^{-,m}_\ell] \},\nonumber\\
\varepsilon_k^{-+} = \varepsilon_k^{+-} =\frac{\xi_k^{-|+} + \xi_k^{+|-}}{2} = \frac{1}{2r}[\Omega^\ell_0\,(U^{-,m}_\ell + U^{+,m}_\ell) - 2U^{0,m}_\ell],\nonumber\\
\varepsilon_k^{00} = \xi_k^{0|0} = \dU^{0,m}_\ell,\nonumber\\
\varepsilon_k^{0+} = \varepsilon^{+0} = \frac{\xi_k^{0|+} + \xi_k^{+|0}}{2} = \frac{1}{2}\{\dU^{+,m}_\ell + \frac{1}{r}[\Omega^\ell_{0}\,U^{0,m}_\ell - U^{+,m}_\ell]\},\nonumber\\
\varepsilon_k^{++} = \xi_k^{+|+} = \frac{1}{r}\Omega^\ell_{2}\,U^{+,m}_\ell(r).\label{impexpr}
\end{eqnarray}
Because $U^- = U^+$ (see Eq.~[\ref{eigexpr}] of Appendix~\ref{gentoreal}), we may simplify these equations to obtain % = \frac{1}{2}\left(\dU^0 - \frac{\Delta_k}{r}\right)
\begin{eqnarray}
\varepsilon_k^{-+} = \varepsilon_k^{+-} = \frac{1}{r}(\Omega^\ell_0 U^{+,m}_\ell - U^{0,m}_\ell),\nonumber\\
\varepsilon_k^{++} = \varepsilon_k^{--}, \,\,\,\,\,\,\,\,\,\,\,\,\varepsilon_k^{0+} = \varepsilon_k^{+0} = \varepsilon_k^{0-} = \varepsilon_k^{-0},\nonumber\\
T_k^{0+} = T_k^{0-},\,\,\,\,\,T_k^{-0} = T_k^{+0},\,\,\,\,T_k^{++} = T_k^{--} = \varepsilon_k^{++} = \varepsilon_k^{--},\,\,\,\,T_k^{00} = \varepsilon_k^{00},\nonumber\\
T_k^{+-} = T_k^{-+}=\varepsilon_k^{+-},\,\,\,\,\,\, Tr(\varepsilon_k) = T_k^{00} - 2T_k^{+-} = \dU^0  + \frac{2}{r}(U^0-\Omega^\ell_0 U^+).
\end{eqnarray}
%The connection between the representation this notation and the original tensor in spherical harmonic coordinates is described in detail in PB73. 
We expand $\bH(\br,\sigma)$ thus
\begin{equation}
\bH(\br,\sigma) = \sum_{s = 0}^\infty\sum_{t=-s}^s h^{ij}_{st}(r,\sigma) Y^{i+j,t}_s\, \be_i\, \be_j,
\end{equation}
where $h^{ij}_{st}$ is the $(i,j)$ component of the tensor $\bH$, and  $i$, $j$ take on values $-1, +1$ or  $0$. We list the components,
\begin{eqnarray}
\siH^{++} = \sum_{s = 0}^\infty\sum_{t=-s}^s h^{++}_{st} Y^{2t}_s,\,\,\,\,\,\,\,\,
\siH^{+-} = \sum_{s = 0}^\infty\sum_{t=-s}^s h^{+-}_{st} Y^{0t}_s,\nonumber\\
\siH^{+0} = \sum_{s = 0}^\infty\sum_{t=-s}^s h^{+0}_{st} Y^{1t}_s,\,\,\,\,\,\,\,\,
\siH^{-0} = \sum_{s = 0}^\infty\sum_{t=-s}^s h^{0-}_{st} Y^{-1t}_s,\nonumber\\
\siH^{00} = \sum_{s = 0}^\infty\sum_{t=-s}^s h^{00}_{st} Y^{0t}_s,\,\,\,\,\,\,\,\,
\siH^{--} = \sum_{s = 0}^\infty\sum_{t=-s}^s h^{--}_{st} Y^{-2t}_s.
\end{eqnarray}

\section{Deriving sensitivity kernels}\label{derive.kernels}
%The double tensor contraction $\bH:{\boldsymbol\mW}$, where $\mW$ was defined in equation~(\ref{defw}), produces the following 
%\begin{equation}
%\int_\odot d\br\, \bH : \mW = \sum_{s,t}\int_\odot d\br\,h_{st}^{00} W^{00} + h_{st}^{--} W^{++} + h_{st}^{++} W^{--} + 2h_{st}^{+-} W^{-+} - 2h_{st}^{0+} W^{0-} - 2h_{st}^{0-} W^{0+}.
%\end{equation}
%Consider the tensor $\mW$, whose components are given by 
%\sum_{\ell=0}^\infty \sum_{m=-\ell}^\ell \sum_{\alpha,\beta} W^{\alpha\beta}\be_\alpha\be_\beta,\\
%&&W^{\alpha\beta} = e_{\alpha\beta}  \sum_{\mu,\gamma}
We obtain the coupling integral in equation~(\ref{coupintfinal}) thus
\begin{eqnarray}
&&\Lambda^{k'}_k= \sum_{s,t} \int_\odot d\br\,\sum_{\alpha, \beta} h_{st}^{\alpha\beta}\,Y^{\alpha+\beta,t}_s \left\{ 
\frac{e_{\alpha\beta}}{2}Tr(\varepsilon^*_{k'})\,Tr(\varepsilon_{k}) (Y^{0m'}_{\ell'})^*\,   \,Y^{0m}_\ell  \right.\nonumber\\
&&\left.
- \left[e_{\alpha\gamma}\,e_{\beta\delta}\,\varepsilon^{\gamma\delta}_k\,Tr(\varepsilon^*_{k'}) (Y^{0m'}_{\ell'})^*\,Y^{\gamma+\delta,m}_\ell + \varepsilon^{\alpha\beta*}_{k'}\,Tr(\varepsilon_{k}) (Y^{\alpha+\beta,m'}_{\ell'})^*\,Y^{0m}_{\ell}\right]
 \right.   \nonumber\\
&&\left. + \sum_{\mu,\gamma} \left[e_{\mu\beta}\, T^{\mu\gamma}_k\,\varepsilon^{\gamma\alpha*}_{k'}\,(Y^{\gamma+\alpha,m'}_{\ell'})^*\,Y^{\gamma+\mu,m}_{\ell}
 + e_{\alpha\mu}\, T^{\mu\gamma}_k\,\varepsilon^{\gamma\beta*}_{k'} \,
(Y^{\gamma+\beta,m'}_{\ell'})^* Y^{\gamma+\mu,m}_{\ell} \right]  \right\},
%\varepsilon^{\alpha\gamma*}_{k'} T^{\beta\mu}_k e_{\mu\gamma}\,Y^{\beta+\mu,m}_\ell (Y^{\alpha+\gamma,m'}_{\ell'})^*  - \varepsilon_{k'}^{\alpha\beta*} Tr(\varepsilon_k)\,Y^{0m}_\ell (Y^{\alpha+\beta,m'}_{\ell'})^*  \nonumber\\
%&&  - \varepsilon_{k}^{\alpha\beta}\, Tr(\varepsilon^*_{k'})\,Y^{\alpha+\beta,m}_\ell (Y^{0m'}_{\ell'})^*,
\end{eqnarray}
where we denote the dot product $\be_\alpha\cdot\be_\beta = e_{\alpha\beta}$.
From the definitions of the unit vectors for generalized coordinates in equation~(\ref{unitvec}), $e_{\alpha\beta} = 0$ with the exceptions $e_{00} =1$ and $e_{+-} = e_{-+} = -1$. The expression resolves into the following 1D problem 
\begin{equation}
\Lambda^{k'}_k= \sum_{s,t}\,\int_\odot dr\, h_{st}^{++}\,\mB_{st}^{++}  \,+\, h_{st}^{00}\,\mB_{st}^{00} \,+\, h_{st}^{--}\,\mB_{st}^{--} \,+\, (h_{st}^{+0} + h_{st}^{0+})\,\mB_{st}^{+0} \,+\, (h_{st}^{-0} + h_{st}^{0-})\,\mB_{st}^{-0}
\, +\, (h_{st}^{-+} + h_{st}^{+-})\,\mB_{st}^{-+},
\end{equation}
where we acknowledge the symmetry of the tensor $\bH$.

%Since $W$ is a symmetric tensor, and therefore only has 6 independent components, we obtain
%\begin{eqnarray}
%W^{++} = 2\varepsilon_{k'}^{0+*}T_{k}^{+0} (Y^{1m'}_{\ell'})^*Y^{1m}_\ell - [Tr(\varepsilon^*_{k'}) \varepsilon_{k}^{++}+ 2\varepsilon_{k'}^{-+*}T_{k}^{++} ] (Y^{0m'}_{\ell'})^* Y^{2m}_\ell \nonumber\\
%- \varepsilon_{k'}^{++*}[2T_{k}^{+-}  + Tr(\varepsilon_k)] (Y^{2m'}_{\ell'})^* Y^{0m}_\ell.
%%= P_{11}\, (Y^{1m'}_{\ell'})^*Y^{1m}_\ell + P_{02}\,(Y^{0m'}_{\ell'})^* Y^{2m}_\ell  + P_{20}\,(Y^{2m'}_{\ell'})^* Y^{0m}_\ell ,
%\end{eqnarray}
Owing to the Wigner addition rules (see Appendix~\ref{symbs}), and because $h_{st}^{++}$ is attached to the harmonic $Y^{2,t}_s$, we have the following expression for $\mB^{++}_{st}$, 
%(THERE MAYBE A MISTAKE HERE MATHEMATICA SAYS $T^{-0}_k$)
\begin{eqnarray}
&&\mB_{st}^{++} = 4\pi\,(-1)^{m'}\,\begin{pmatrix}\ell' & s & \ell \\ -m' & t & m\end{pmatrix} \left[-\varepsilon^{++}_k\, \varepsilon^{00*}_{k'}\, \begin{pmatrix}\ell' & s & \ell \\ 0 & 2 & -2\end{pmatrix} 
+ 2 T^{-0}_k\, \varepsilon^{0+*}_{k'}\, \begin{pmatrix}\ell' & s & \ell \\ -1 & 2 & -1\end{pmatrix} \right.\nonumber\\
&&\left.- \varepsilon^{00}_k\,\varepsilon^{++*}_{k'} \begin{pmatrix}\ell' & s & \ell \\ -2 & 2 & 0\end{pmatrix} \right].\label{bpp}
%&&\int_\odot d\br\, W^{++} h_{st}^{--} = -\int_\odot d\br\, h_{st}^{--} [Tr(\varepsilon^*_{k'}) \varepsilon_{k}^{++}+ 2\varepsilon_{k'}^{-+*}T_{k}^{++} ] (Y^{0m'}_{\ell'})^* Y^{-2,s}_tY^{2m}_\ell\, \nonumber\\
%&&= -4\pi(-1)^{m'}\begin{pmatrix}\ell' & s & \ell \\ -m' & t & m\end{pmatrix}\begin{pmatrix}\ell' & s & \ell \\ 0 & -2 & 2\end{pmatrix}\int_\odot dr\, r^2 h_{st}^{--} \varepsilon^{++}\varepsilon^{00'*}.
\end{eqnarray}

A similar analysis may be applied to obtain the kernel for $h_{st}^{--}$,
\begin{eqnarray}
&&\mB_{st}^{--} = 4\pi\,(-1)^{m'}\,\begin{pmatrix}\ell' & s & \ell \\ -m' & t & m\end{pmatrix} \left[-\varepsilon^{++}_k\, \varepsilon^{00*}_{k'}\, \begin{pmatrix}\ell' & s & \ell \\ 0 & -2 & 2\end{pmatrix} 
+ 2T^{+0}_k\, \varepsilon^{0-*}_{k'}\, \begin{pmatrix}\ell' & s & \ell \\ 1 & -2 & 1\end{pmatrix} \right.\nonumber\\
&&\left.- \varepsilon^{00}_k\,\varepsilon^{++*}_{k'} \begin{pmatrix}\ell' & s & \ell \\ 2 & -2 & 0\end{pmatrix} \right]
= (-1)^{\ell'+\ell + s }\, \mB_{st}^{++},\label{bmm}
\end{eqnarray}
where we have used the degeneracy of the $\pm$ components of the eigenfunction (see appendix~\ref{gentoreal}).
%We may simplify this further using the $B$ symbols (Appendix~\ref{symbs}),
%\begin{equation}
%\begin{pmatrix}\ell' & s & \ell \\ 0 & -2 & 2\end{pmatrix} = (-1)^{\ell'+s+\ell}\begin{pmatrix}s & \ell' & \ell \\ -2 & 0 & 2\end{pmatrix} =
%\frac{1+(-1)^{\ell'+s+\ell}}{2}\begin{pmatrix}s & \ell' & \ell \\ -2 & 0 & 2\end{pmatrix}
%= \frac{B^{(2)+}_{s\ell'\ell}}{4\Omega^s_0\Omega^s_2\Omega^{\ell}_0\Omega^\ell_2}\label{sym2},
%\end{equation}
%where we note from equations (C.221) and (C.222) of \citet{DT98} that the Wigner symbol in equation~(\ref{sym2}) is non-zero only for even $\ell'+\ell +s$ (see Eq.~[\ref{relate1}]), thereby allowing us to constructing this relationship.
%The kernel for the $h_{st}^{++}$ Lorentz-stress contribution is
%\begin{eqnarray}
% \mB_{st}^{--} = -4\pi r(-1)^{m'} \Omega^\ell_2 \dU^{0'*} U^+\begin{pmatrix}\ell' & s & \ell \\ 0 & -2 & 2\end{pmatrix}\begin{pmatrix}\ell' & s & \ell \\ -m' & t & m\end{pmatrix}.
%\end{eqnarray}

Next we compute the kernel for the $h_{st}^{0+} + h_{st}^{+0}$ component, 
%IT APPEARS THERE IS $T_k^{0-}\varepsilon_{k'}^{+-*} - T^{-0}_k\varepsilon_{k'}^{00*}]$  AND THERE IS A $T_k^{0+}$
%\begin{eqnarray}
%&&\mB_{st}^{0+} = \mB_{st}^{+0} = 4\pi\,(-1)^{m'}\,\begin{pmatrix}\ell' & s & \ell \\ -m' & t & m\end{pmatrix}\,\left\{ [\varepsilon^{0-}_{k}\,Tr(\varepsilon^*_{k'}) + T^{0-}_k\,\varepsilon_{k'}^{+-*} - T^{-0}_k\,\varepsilon_{k'}^{00*}] \begin{pmatrix}\ell' & s & \ell \\ 0 & 1 & -1\end{pmatrix} \right.\nonumber\\
%&&\left. - [\varepsilon^{0+*}_{k}\,Tr(\varepsilon_{k}) + \varepsilon^{+-}_k\,\varepsilon_{k'}^{+0*} - \varepsilon_k^{00}\,\varepsilon_{k'}^{0+*}]\begin{pmatrix}\ell' & s & \ell \\ -1 & 1 & 0\end{pmatrix} 
%+ \varepsilon_k^{--}\varepsilon^{-0*}_{k'}
%\begin{pmatrix}\ell' & s & \ell \\ 1 & 1 & -2\end{pmatrix} +  T_k^{0+}\varepsilon^{++*}_{k'} \begin{pmatrix}\ell' & s & \ell \\ -2 & 1 & 1\end{pmatrix}
%\right\}.\label{b0p}
%\end{eqnarray}
%
%This may be simplified further,
\begin{eqnarray}
&&2\mB_{st}^{0+} = 2\mB_{st}^{+0} = 4\pi\,(-1)^{m'}\,\begin{pmatrix}\ell' & s & \ell \\ -m' & t & m\end{pmatrix}\,\left\{ [(T^{0+}_{k}- T^{+0}_k)\,\varepsilon_{k'}^{00*} - 2T^{+0}_k\,\varepsilon_{k'}^{+-*}] \begin{pmatrix}\ell' & s & \ell \\ 0 & 1 & -1\end{pmatrix} \right.\nonumber\\
&&\left. - 2\varepsilon^{+-}_k\,\varepsilon^{0+*}_{k}\,\begin{pmatrix}\ell' & s & \ell \\ -1 & 1 & 0\end{pmatrix} 
+ 2\varepsilon_k^{--}\varepsilon^{-0*}_{k'}
\begin{pmatrix}\ell' & s & \ell \\ 1 & 1 & -2\end{pmatrix} +  2T_k^{0+}\varepsilon^{++*}_{k'} \begin{pmatrix}\ell' & s & \ell \\ -2 & 1 & 1\end{pmatrix}
\right\}.\label{b0p}
\end{eqnarray}

The symmetries between the $\pm$ terms encourage us to consider $h_{st}^{0-} + h_{st}^{-0}$ next, and we obtain,
%IT APPEARS THERE IS $T_k^{0+}\varepsilon_{k'}^{+-*} - T^{+0}_k\varepsilon_{k'}^{00*}]$  AND THERE IS A $T_k^{0-}$
%\begin{eqnarray}
%&&\mB_{st}^{0-} = 4\pi\,(-1)^{m'}\,\begin{pmatrix}\ell' & s & \ell \\ -m' & t & m\end{pmatrix}\,\left\{\varepsilon_k^{++}\varepsilon^{+0*}_{k'}
%\begin{pmatrix}\ell' & s & \ell \\ -1 & -1 & 2\end{pmatrix} + T^{+0}_k [\varepsilon_{k'}^{-+*} - \varepsilon_{k'}^{00*}] \begin{pmatrix}\ell' & s & \ell \\ 0 & -1 & 1\end{pmatrix} \right.\nonumber\\
%&&\left. + \varepsilon^{-0*}_{k'} [\varepsilon_k^{+-} - \varepsilon_k^{00}] \begin{pmatrix}\ell' & s & \ell \\ 1 & -1 & 0\end{pmatrix}
\begin{equation}
%+ T_k^{-0}\varepsilon^{--*}_{k'} \begin{pmatrix}\ell' & s & \ell \\ 2 & -1 & -1\end{pmatrix}
%\right\} 
\mB_{st}^{0-} = \mB_{st}^{-0} = (-1)^{\ell'+\ell + s }\, \mB_{st}^{0+}.\label{b0m}
\end{equation}
%where we have exploited the degeneracy between $\pm$ components of eigenfunctions and their derivatives.

The penultimate term is $h_{st}^{00}$ whose kernel is 
%(JISHNU IS GETTING A FACTOR OF 3 ON $Tr(\varepsilon_k)\,\varepsilon^{00*}_{k'}$.
\begin{eqnarray}
&&\mB_{st}^{00} = 4\pi\,(-1)^{m'}\,\frac{1+(-1)^{\ell'+\ell + s }}{2}\begin{pmatrix}\ell' & s & \ell \\ -m' & t & m\end{pmatrix}\,\left\{
-4T_k^{0-}\,\varepsilon_{k'}^{-0*}\, \begin{pmatrix}\ell' & s & \ell \\ -1 & 0 & 1\end{pmatrix} \right. \nonumber\\
&&\left.  + \frac{(2\varepsilon^{+-}_{k} + \varepsilon^{00}_{k})(2\varepsilon^{+-*}_{k'} + \varepsilon^{00*}_{k'}) }{2} \begin{pmatrix}\ell' & s & \ell \\ 0 & 0 & 0\end{pmatrix}
\right\}.\label{b00}
\end{eqnarray}

Finally, we have the expression for the kernel for $h_{st}^{+-} + h_{st}^{-+}$,
\begin{eqnarray}
&&2\mB_{st}^{+-} = 4\pi\,(-1)^{m'}\,\begin{pmatrix}\ell' & s & \ell \\ -m' & t & m\end{pmatrix}\,\frac{1+(-1)^{\ell'+\ell + s }}{2}\,\left\{
4T_k^{+0}\,\varepsilon_{k'}^{-0*}\, \begin{pmatrix}\ell' & s & \ell \\ -1 & 0 & 1\end{pmatrix}
 \right. \nonumber\\
&&\left. - 4\varepsilon_k^{++}\,\varepsilon_{k'}^{++*}\, \begin{pmatrix}\ell' & s & \ell \\ -2 & 0 & 2\end{pmatrix} - \varepsilon^{00*}_{k'}\varepsilon^{00}_{k} \begin{pmatrix}\ell' & s & \ell \\ 0 & 0 & 0\end{pmatrix}
\right\}  = 2\mB_{st}^{-+},\label{bpm}
\end{eqnarray}
%\left[\frac{Tr(\varepsilon^*_{k'})\,Tr(\varepsilon_{k})}{2} + 2\varepsilon^{+-*}_{k'}\varepsilon^{+-}_{k} + \varepsilon^{+-*}_{k'}\,Tr(\varepsilon_k) + \varepsilon^{+-}_{k}\, Tr(\varepsilon^*_{k'}) \right]
where we have exploited the symmetric nature of the Lorentz stress tensor. The structure of these kernels allows for rewriting the inverse problem thus,
\begin{equation}
\Lambda^{k'}_k= \sum_{s,t}\,\int_\odot dr\,\, \mB_{st}^{++}  \,[h_{st}^{++} + (-1)^{\ell'+\ell+s}\,h_{st}^{--}]\,+\, \mB_{st}^{00} \,h_{st}^{00}\,+\, 2\,\mB_{st}^{+0}\,[h_{st}^{+0} + (-1)^{\ell'+\ell+s}\,h_{st}^{0-}]\,+\, 2\,\mB_{st}^{-+}\,h_{st}^{-+},\label{final.int}
\end{equation}
and is therefore sensitive to the sums or differences between various components of $h$ depending on whether the sum $\ell' + \ell + s$ is even or odd. Given the complexity of these expressions, we additionally verified the kernels using Mathematica.
\section{List of Symbols}\label{symbs}
\begin{eqnarray}
\gamma_\ell = \sqrt{\frac{2\ell+1}{4\pi}},\\
\Omega^\ell_N = \sqrt{\frac{1}{2}(\ell+N)(\ell-N+1)},\label{omegdef}\\
\Omega^\ell_0 = \Omega^\ell_1,\,\,\,\,\,\,\,\,\,\,\,\,\,\, \Omega^\ell_{-1} = \Omega^\ell_2.\label{omegequal}
\end{eqnarray}

The definition of the Wigner-3$j$ symbol is
\begin{eqnarray}
\int_0^{2\pi} d\phi \int_0^{\pi} d\theta\,\sin\theta (Y^{N'm'}_{\ell'})^* Y^{N{''}m{''}}_{\ell{''}} Y^{Nm}_\ell = \nonumber\\
4\pi(-1)^{(N'-m')} \begin{pmatrix} \ell' & \ell{''} & \ell \\ -N' & N{''} & N\end{pmatrix} \begin{pmatrix}\ell' & \ell{''} & \ell \\ -m' & m{''} & m\end{pmatrix}.
\end{eqnarray}
Each Wigner symbol is non-zero only if the elements in the second row sum to zero, i.e. $N{''} + N - N' =0$ and $m{''} + m - m' =0$.
We also use
\begin{equation}
\begin{pmatrix} \ell' & \ell{''} & \ell \\ N' & -N{''} & -N\end{pmatrix} = (-1)^{\ell'+\ell+s}\begin{pmatrix} \ell' & \ell{''} & \ell \\ -N' & N{''} & N\end{pmatrix}.
\end{equation}
%The following properties are of use in the current analysis
%\begin{eqnarray}
%\begin{pmatrix}\ell' & s & \ell \\ -m' & t & m\end{pmatrix} = \begin{pmatrix}s & \ell & \ell' \\ t & m & -m'\end{pmatrix},\label{rel1}\\
%\begin{pmatrix}\ell' & s & \ell \\ -m' & t & m\end{pmatrix} = (-1)^{\ell+s+\ell'}\begin{pmatrix}s & \ell & \ell' \\ -t & -m & m'\end{pmatrix},
%\label{rel2}
%\end{eqnarray}
%where $m'=m+t$ in equations~(\ref{rel1}) and~(\ref{rel2}) in order that the symbols be non-zero.
%The $B$-coefficient symbol, e.g. \citet{woodhouse80, lavely92,DT98}, simplifies expressions so we use it here,
%\begin{equation}
%B^{(N)\pm}_{s\ell'\ell } = \frac{(-1)^N}{2}(1\pm(-1)^{\ell+\ell'+s})\left[\frac{(s+N)!(\ell+N)!}{(s-N)!(\ell-N)!}\right]^{1/2} \begin{pmatrix} s & \ell' & \ell \\ -N & 0 & N\end{pmatrix}.
%\end{equation}
%We also have
%\begin{equation}
%\begin{pmatrix} s & \ell' & \ell \\ 0 & 0 & 0\end{pmatrix} = \frac{1+(-1)^{\ell+\ell'+s}}{2}\begin{pmatrix} s & \ell' & \ell \\ 0 & 0 & 0\end{pmatrix},\label{relate0}
%\end{equation}
%because this symbol is nonzero only for even $\ell+\ell'+s$.
%From equations~(C221) and~(C222), the following relationships hold, 
%\begin{equation}
%\begin{pmatrix} s & \ell' & \ell \\ -2 & 0 & 2\end{pmatrix} = \frac{1+(-1)^{\ell+\ell'+s}}{2}\begin{pmatrix} s & \ell' & \ell \\ -2 & 0 & 2\end{pmatrix},\label{relate2}
%\end{equation}
%and
%\begin{equation}
%\begin{pmatrix} s & \ell' & \ell \\ -1 & 0 & 1\end{pmatrix} = \frac{1+(-1)^{\ell+\ell'+s}}{2}\begin{pmatrix} s & \ell' & \ell \\ -1 & 0 & 1\end{pmatrix}.\label{relate1}
%\end{equation}
\section{Converting from the generalized to spherical coordinates}\label{gentoreal}
The eigenfunctions of a spherically symmetric model $U,V$ in generalized coordinates are
\begin{equation}
\xi_k^0 = \gamma_\ell\, U,\,\,\,\,\,\,\,\,\,\,\, \xi_k^+ = \xi_k^- = \gamma_\ell\, \Omega^\ell_0\, V.\label{eigexpr}
\end{equation}

We have the following relations between the $\pm,0$ vectors to the $(r,\theta,\phi)$ representation,
\begin{eqnarray}
%&&u^t_s = \frac{B^{0,t}_s}{\gamma_s},\,\,\,\,\,\,\, v^t_s = \frac{B^{-,t}_s + B^{+,t}_s}{2\gamma_s\Omega^s_0},\,\,\,\,\,\,\, w^t_s = \frac{i}{2\gamma_s\Omega^s_0}(B^{-,t}_s - B^{+,t}_s),\\
&&\be_0 = \rhat,\,\,\,\,\be_- = \frac{\be_\theta - i\be_\phi}{\sqrt2},\,\,\,\,\be_+ = -\frac{\be_\theta + i\be_\phi}{\sqrt2}.
%&&B^{0,t}_s = \gamma_s B^t_{s,r},\,\,\,\,\,\,\,  \frac{B^{-,t}_s - B^{+,t}_s}{\sqrt{2}} = \gamma_s B^t_{s,\theta},\,\,\,\,\,\,\, -i\frac{B^{-,t}_s + B^{+,t}_s}{\sqrt{2}} = \gamma_s B^t_{s,\phi}.
%&&\langle B^t_{s,r} B^t_{s,r} \rangle = \frac{h^{00,t}_s}{\gamma_s},\,\,\,\,  \langle B^t_{s,\theta} B^t_{s,\theta} \rangle = \frac{h^{--,t}_s + h^{++,t}_s}{2\gamma_s} + \frac{h^{+-,t}_s}{\gamma_s},\nonumber\\
%&&\langle B^t_{s,\phi} B^t_{s,\phi} \rangle = -\frac{h^{--,t}_s + h^{++,t}_s}{2\gamma_s} + \frac{h^{+-,t}_s}{\gamma_s}.\\
%&&\langle u^t_s\,u^t_s\rangle = \frac{h^{00}}{\gamma_s\gamma_s},\,\,\,\,\,\,\, \langle u^t_s\,v^t_s\rangle = \frac{h^{0-} + h^{0+}}{2\gamma_s\gamma_s\Omega^s_0},\,\,\,\,\,\,\,\langle u^t_s\,w^t_s\rangle = i\frac{h^{0-} - h^{0+}}{2\gamma_s\gamma_s\Omega^s_0},\,\,\,\,\,\,\,\\
%&&\langle v^t_s\,v^t_s\rangle = \frac{h^{--} + h^{++} +2h^{+-}}{4\gamma_s\gamma_s\Omega^s_0\Omega^s_0},\,\,\,\,\,\,\, \langle w^t_s\,v^t_s\rangle = i\frac{h^{--} - h^{++}}{4\gamma_s\gamma_s\Omega^s_0\Omega^s_0},\,\,\,\,\,\,\,\langle w^t_s\,w^t_s\rangle = \frac{2h^{+-} - h^{--} - h^{++}}{4\gamma_s\gamma_s\Omega^s_0\Omega^s_0}.\nonumber\\
\end{eqnarray}
To reconstruct the real-space version of $\bH$, we first note that
\begin{eqnarray}
\bH = \sum_{s,t} \left[h_{st}^{++} \be_+ \be_+ Y^{2,t}_s + (h_{st}^{0+} \be_0 \be_+ + h_{st}^{+0} \be_+ \be_0) Y^{1,t}_s \right.\nonumber\\
\left. + (h_{st}^{00} \be_0 \be_0  + h_{st}^{+-} \be_+ \be_- + h_{st}^{-+} \be_- \be_+) Y^{0,t}_s\right.\nonumber\\
\left. +  (h_{st}^{0-} \be_0 \be_- + h_{st}^{-0} \be_- \be_0) Y^{-1,t}_s+h_{st}^{--}\be_- \be_- Y^{-2,t}_s\right].
\end{eqnarray}
The components of $\bH$ are obtained by dotting with $(\rhat, \be_\theta , \be_\phi)$. We compute the following,
\begin{eqnarray}
\rhat \cdot \be_- = 0,\,\,\,\, \be_\theta \cdot \be_- = \frac{1}{\sqrt2},\,\,\,\, \be_\phi \cdot \be_- = -\frac{i}{\sqrt2},\nonumber\\
\rhat \cdot \be_0 = 1,\,\,\,\, \be_\theta \cdot \be_0 = 0,\,\,\,\, \be_\phi \cdot \be_0 = 0,\nonumber\\
\rhat \cdot \be_+ = 0,\,\,\,\, \be_\theta \cdot \be_+ = -\frac{1}{\sqrt2},\,\,\,\, \be_\phi \cdot \be_+ = -\frac{i}{\sqrt2}.
\end{eqnarray}

Because $\bH$ is symmetric, we only need six components,
\begin{eqnarray}
B_r B_r = \rhat \rhat:\bH = \sum_{s,t} h_{st}^{00} Y^{0,t}_s ,\\
B_r B_\theta = \rhat\, \be_\theta:\bH = \frac{1}{\sqrt2}\left[\sum_{s,t} h_{st}^{0-} Y^{-1,t}_s - h_{st}^{0+} Y^{1,t}_s\right],\\
B_r B_\phi = \rhat\, \be_\phi:\bH = -\frac{i}{\sqrt2}\left[\sum_{s,t} h_{st}^{0-} Y^{-1,t}_s + h_{st}^{0+} Y^{1,t}_s\right],\\
B_\theta B_\theta = \be_\theta\, \be_\theta:\bH = \frac{1}{2}\left[\sum_{s,t} h_{st}^{++} Y^{2,t}_s -2h_{st}^{+-} Y^{0,t}_s + h_{st}^{--} Y^{-2,t}_s\right],\\
B_\theta B_\phi = \be_\theta\, \be_\phi:\bH = \frac{i}{2}\left[\sum_{s,t} h_{st}^{++} Y^{2,t}_s - h_{st}^{--} Y^{-2,t}_s\right],\\
B_\phi B_\phi = \be_\phi\, \be_\phi:\bH = -\frac{1}{2}\left[\sum_{s,t} h_{st}^{++} Y^{2,t}_s  + 2h_{st}^{+-} Y^{0,t}_s + h_{st}^{--} Y^{-2,t}_s\right].
\end{eqnarray}
%Note that we only have access to $h^{00}$, $h^{+-}$, $h^{++} + h^{--}$ and $h^{0+} + h^{0-}$, which implies that only $\langle B_r B_r \rangle$ and $\langle B_\theta B_\theta + B_\phi B_\phi \rangle$ are fully known. All the other components are known partially at best,
We rewrite the above equations in terms of sums and differences in the $h_{st}$
\begin{eqnarray}
&& B_r B_r  = \rhat \rhat:\bH = \sum_{s,t} h_{st}^{00} Y^{0,t}_s ,\\
&& B_r B_\theta  = \frac{1}{\sqrt2}\left[\sum_{s,t} \left[(h_{st}^{0-} + h_{st}^{0+})\frac{Y^{-1,t}_s -Y^{1,t}_s}{2}\right] + (h_{st}^{0-}- h_{st}^{0+})\frac{Y^{-1,t}_s + Y^{1,t}_s}{2}\right],\\
&& B_r B_\phi  = -\frac{i}{\sqrt2}\left[\sum_{s,t} \left[(h_{st}^{0-} + h_{st}^{0+})\frac{Y^{-1,t}_s +Y^{1,t}_s}{2}\right] + (h_{st}^{0-}- h_{st}^{0+})\frac{Y^{-1,t}_s - Y^{1,t}_s}{2}\right],\\
&& B_\theta B_\theta  =\frac{1}{2}\left[\sum_{s,t} \left[(h_{st}^{--} + h_{st}^{++})\frac{Y^{-2,t}_s+Y^{2,t}_s}{2} -2h_{st}^{+-} Y^{0,t}_s\right] + (h_{st}^{--} - h_{st}^{++})\frac{Y^{-2,t}_s - Y^{2,t}_s}{2}\right],\nonumber\\
&& B_\theta B_\phi  = -\frac{i}{2}\left[\sum_{s,t} \left[(h_{st}^{--} + h_{st}^{++})\frac{Y^{-2,t}_s - Y^{2,t}_s}{2}\right] + (h_{st}^{--} - h_{st}^{++})\frac{Y^{-2,t}_s + Y^{2,t}_s}{2}\right],\\
&& B_\phi B_\phi  =-\frac{1}{2}\left[\sum_{s,t} \left[(h_{st}^{--} + h_{st}^{++})\frac{Y^{-2,t}_s+Y^{2,t}_s}{2} +2h_{st}^{+-} Y^{0,t}_s\right] + (h_{st}^{--} - h_{st}^{++})\frac{Y^{-2,t}_s - Y^{2,t}_s}{2}\right].\nonumber\\
\end{eqnarray}
Because of the $1 + (-1)^{\ell' + \ell + s}$ multiplying factor in equations~(\ref{b00}) and~(\ref{bpm}), the components $h_{st}^{+-}$ and $h_{st}^{00}$ are only sensed by modes when the sum $\ell + \ell' + s$ is even.
Similarly, the even or odd parity of $\ell' + \ell + s$ determines whether we are able to infer sums or differences, i.e. $h_{st}^{--}\, (-1)^{\ell'+\ell+s} + h_{st}^{++}$ and $h_{st}^{0-}\,(-1)^{\ell'+\ell+s} + h_{st}^{0+}$ respectively (Eqs.~[\ref{bpp}] through~[\ref{b0p}]). This effect is akin to being able to infer toroidal flows only when $\ell' + \ell + s$ is odd and poloidal flows when $\ell' + \ell + s$ is even \citep[e.g., Appendices C and D of][]{lavely92} and Appendix~\ref{flowkern}.

\section{Deriving Flow Kernels}\label{flowkern}
We sketch the technique to compute kernels for flows using generalized coordinates. Indeed, \citet{lavely92} discuss this possibility in their Appendix C but do not pursue it. We begin by expressing a general flow field $\bu_0$ thus
\begin{equation}
\bu_0 = \sum_{s=0}^\infty\sum_{t=-s}^s u^+_{st}\, Y^{1,t}_s\, \be_+ + u^0_{st}\, Y^{0,t}_s\, \be_0
+u^-_{st}\, Y^{-1,s}_t\, \be_-.\label{vecflow}
\end{equation}
The relationship between the $\pm,0$ symbols and poloidal and toroidal flow components is \citep{phinney73}
\begin{equation}
u^t_s = \frac{u^{0}_{st}}{\gamma_s},\,\,\,\,\,\,\, v^t_s = \frac{u^{-}_{st} + u^{+}_{st}}{2\gamma_s\Omega^s_0},\,\,\,\,\,\,\, w^t_s = \frac{i}{2\gamma_s\Omega^s_0}(u^{-}_{st} - u^{+}_{st}),\label{poltor}
\end{equation}
where $u^t_s, v^t_s$ represent the poloidal flow components and $w^t_s$ is the toroidal flow component. The perturbation to the wave operator~(\ref{fullop}) due to advection is given by
\begin{equation}
\delta\op\bxi = -2i\omega\,\rho\bu_0\cdot\bnabla\bxi,
\end{equation}
and recalling equation~(\ref{coupmat}), the coupling between two modes $k$ and $k'$ induced by flows is given by
\begin{equation}
\Lambda^{k'}_k = 2i\omega\,\int_\odot d\br\, \rho\bu_0\cdot(\bnabla\bs_{k})\cdot\bs^*_{k'}.
\end{equation}
In generalized-coordinate notation, this becomes
\begin{eqnarray}
&&\Lambda^{k'}_k = 2i\omega\,\int_\odot d\br\, 
[- \rho\,u^+_{st}\, T^{--}_{k}\, \xi^{-*}_{k'}\, (Y^{-1m'}_{\ell'})^* Y^{1t}_s Y^{-2m}_\ell 
- \rho\, u^+_{st}\, T^{-0}_{k}\, \xi^{0*}_{k'}\, (Y^{0m'}_{\ell'})^* Y^{1t}_s Y^{-1m}_\ell \nonumber\\
&&- \rho\, u^+_{st}\, T^{-+}_{k}\, \xi^{+*}_{k'}\, (Y^{1m'}_{\ell'})^* Y^{1t}_s Y^{0m}_\ell
+ \rho\, u^0_{st}\, T^{0-}_{k}\, \xi^{-*}_{k'}\, (Y^{-1m'}_{\ell'})^* Y^{0t}_s Y^{-1m}_\ell \nonumber\\
&&+ \rho\, u^0_{st}\, T^{00}_{k}\, \xi^{0*}_{k'}\, (Y^{0m'}_{\ell'})^* Y^{0t}_s Y^{0m}_\ell
+ \rho\, u^0_{st}\, T^{0+}_{k}\, \xi^{+*}_{k'}\, (Y^{1m'}_{\ell'})^* Y^{0t}_s Y^{1m}_\ell \nonumber\\
&&- \rho\, u^-_{st}\, T^{+-}_{k}\, \xi^{-*}_{k'}\, (Y^{-1m'}_{\ell'})^* Y^{-1t}_s Y^{0m}_\ell
- \rho\, u^-_{st}\, T^{+0}_{k}\, \xi^{0*}_{k'}\, (Y^{0m'}_{\ell'})^* Y^{-1t}_s Y^{1m}_\ell \nonumber\\
&&-\rho\, u^-_{st}\, T^{++}_{k}\, \xi^{+*}_{k'}\, (Y^{1m'}_{\ell'})^* Y^{-1t}_s Y^{2m}_\ell].
\end{eqnarray}

The spherical integration reduces all these terms to Wigner-3$j$ symbols,
\begin{eqnarray}
&&\Lambda^{k'}_k = 8i\pi \omega\,(-1)^{m'}\int_\odot dr\,\rho\,r^2 \begin{pmatrix}\ell' & s & \ell \\ -m' & t & m \end{pmatrix}\left[ u^+_{st}\, T^{--}_{k}\, \xi^{-*}_{k'}\, \begin{pmatrix}\ell' & s & \ell \\ 1 & 1 & -2 \end{pmatrix} \right. \nonumber\\
&&\left. - u^+_{st}\, T^{-0}_{k}\, \xi^{0*}_{k'}\, \begin{pmatrix}\ell' & s & \ell \\ 0 & 1 & -1 \end{pmatrix}
+ u^+_{st}\, T^{-+}_{k}\, \xi^{+*}_{k'}\, \begin{pmatrix}\ell' & s & \ell \\ -1 & 1 & 0 \end{pmatrix}
+ u^-_{st}\, T^{+-}_{k}\, \xi^{-*}_{k'}\, \begin{pmatrix}\ell' & s & \ell \\ 1 & -1 & 0 \end{pmatrix} \right.
\nonumber\\
&&\left. - u^-_{st}\, T^{+0}_{k}\, \xi^{0*}_{k'}\, \begin{pmatrix}\ell' & s & \ell \\ 0 & -1 & 1 \end{pmatrix}
+u^-_{st}\, T^{++}_{k}\, \xi^{+*}_{k'}\, \begin{pmatrix}\ell' & s & \ell \\ -1 & -1 & 2 \end{pmatrix} 
- u^0_{st}\, T^{0-}_{k}\, \xi^{-*}_{k'}\, \begin{pmatrix}\ell' & s & \ell \\ 1 & 0 & -1 \end{pmatrix} \right.
\nonumber\\
&&\left.+ u^0_{st}\, T^{00}_{k}\, \xi^{0*}_{k'}\, \begin{pmatrix}\ell' & s & \ell \\ 0 & 0 & 0 \end{pmatrix}
- u^0_{st}\, T^{0+}_{k}\, \xi^{+*}_{k'}\,  \begin{pmatrix}\ell' & s & \ell \\ -1 & 0 & 1 \end{pmatrix}\right].
\end{eqnarray}
Because of the $\pm$ degeneracy in the eigenfunctions and the corresponding expressions for $T$, i.e. $T^{0\pm} = T^{0\mp}$, $T^{\pm0} = T^{\mp0}$, $T^{++} = T^{--}$ and $T^{+-} = T^{-+}$, this expression may reduced,
\begin{eqnarray}
&&\Lambda^{k'}_k = 8i\pi \omega\,(-1)^{m'}\int_\odot dr\,\rho\,r^2 \begin{pmatrix}\ell' & s & \ell \\ -m' & t & m \end{pmatrix}\left\{ \left[u^+_{st} + (-1)^{\ell'+\ell+s}\,u^-_{st}\right]\left[ T^{--}_{k}\, \xi^{-*}_{k'}\, \begin{pmatrix}\ell' & s & \ell \\ 1 & 1 & -2 \end{pmatrix}\right. \right. \nonumber\\
&&\left.\left. -  T^{-0}_{k}\, \xi^{0*}_{k'}\, \begin{pmatrix}\ell' & s & \ell \\ 0 & 1 & -1 \end{pmatrix}
+  T^{-+}_{k}\, \xi^{+*}_{k'}\, \begin{pmatrix}\ell' & s & \ell \\ -1 & 1 & 0 \end{pmatrix}\right]
%+ u^-_{st}\, T^{+-}_{k}\, \xi^{-*}_{k'}\, \begin{pmatrix}\ell' & s & \ell \\ 1 & -1 & 0 \end{pmatrix} \right.
\right.
\nonumber\\ %[u^+_{st} + (-1)^{\ell'+\ell+s}\,u^-_{st}]\, [u^+_{st} + (-1)^{\ell'+\ell+s}\,u^-_{st}]\,
%&&\left. - u^-_{st}\, T^{+0}_{k}\, \xi^{0*}_{k'}\, \begin{pmatrix}\ell' & s & \ell \\ 0 & 1 & -1 \end{pmatrix}
%+u^-_{st}\, T^{++}_{k}\, \xi^{+*}_{k'}\, \begin{pmatrix}\ell' & s & \ell \\ -1 & -1 & 2 \end{pmatrix} 
%- u^0_{st}\, T^{0-}_{k}\, \xi^{-*}_{k'}\, \begin{pmatrix}\ell' & s & \ell \\ 1 & 0 & -1 \end{pmatrix} \right.
%\right.
%\nonumber\\
&&\left.+ u^0_{st}\,\frac{1+ (-1)^{\ell'+\ell+s}}{2}\left[T^{00}_{k}\, \xi^{0*}_{k'} \begin{pmatrix}\ell' & s & \ell \\ 0 & 0 & 0 \end{pmatrix}
-2 T^{0+}_{k}\, \xi^{+*}_{k'}\,  \begin{pmatrix}\ell' & s & \ell \\ -1 & 0 & 1 \end{pmatrix}\right]\right\}.\label{flowlam}
\end{eqnarray}

%\begin{eqnarray}
%&&\Lambda^{k'}_k = 8i\pi \omega\,(-1)^{m'}\,\int_\odot dr\,\rho\,r^2 \begin{pmatrix}\ell' & s & \ell \\ -m' & t & m \end{pmatrix}\left\{
%u^0_{st}\, T^{00}_{k}\, \xi^{0}_{k'}\, \begin{pmatrix}\ell' & s & \ell \\ 0 & 0 & 0 \end{pmatrix} \right. \nonumber\\
%&&\left.-  \left[u^+_{st} + (-1)^{\ell'+\ell+s} u^-_{st}\right]\, T^{-0}_{k}\, \xi^{0}_{k'}\, \begin{pmatrix}\ell' & s & \ell \\ 0 & 1 & -1 \end{pmatrix}\right\}.\label{flowlam}
%\end{eqnarray}
Equation~(\ref{flowlam}) states that $u^0_{st}$, which from equation~(\ref{poltor}) is directly proportional to the radial flow, can only be inferred for even values of the sum $\ell'+\ell + s$ (the Wigner-3$j$ symbol with all zeros in the second row is zero for odd $\ell'+\ell+s$). Similarly, depending on whether $\ell'+\ell+s$ is even or odd, we correspondingly recover the sum $u^+_{st} + u^-_{st}$ or difference $u^+_{st} - u^-_{st}$, giving us alternate access to the poloidal flow (for even $\ell' + \ell +s$) and toroidal flow (when $\ell' + \ell +s$ is odd). See also \citet{lavely92}, \citet{woodard14} and \citet{hanasoge_etal_2017} for more details.

\end{document}